\documentclass[aps,pra,twocolumn,showpacs,superscriptaddress]{revtex4-1}
\usepackage{amsmath} 
\usepackage{amssymb}
\usepackage{mathtools}
\usepackage{float}
\usepackage{graphicx,graphics,color}
\renewcommand{\vec}[1]{\boldsymbol{#1}}

\begin{document}
\title{Suppression and Control of Pre-thermalization  in Multi-component\\
Fermi Gases Following a Quantum Quench}
\author{Chen-How Huang}
\affiliation{Department of Physics, National Tsing Hua University, Hsinchu 30013, Taiwan}
\author{Yosuke Takasu}
\affiliation{Department of Physics, Graduate School of Science, Kyoto University, Kyoto 606-8502, Japan}
\author{Yoshiro Takahashi}
\affiliation{Department of Physics, Graduate School of Science, Kyoto University, Kyoto 606-8502, Japan}
\author{Miguel A. Cazalilla}
\affiliation{Department of Physics, National Tsing Hua University, Hsinchu 30013, Taiwan}
\affiliation{National Center for Theoretical Sciences (NCTS), Hsinchu City, Taiwan}
\affiliation{Donostia International Physics Center (DIPC), 
Manuel de Lardizabal, 4 San Sebastian, Spain}
\date{\today}
\date{\today}
\begin{abstract}
We investigate the mechanisms of control and suppression  of pre-thermalization in  $N$-component alkaline earth gases. To this end, we compute the short-time dynamics   of the instantaneous momentum distribution  and the relative population  for different initial conditions after an interaction quench,  accounting for the 
effect of initial interactions.   We find that switching on an interaction that breaks the  SU$(N)$ symmetry of the initial Hamiltonian, thus allowing for the occurrence of spin-changing collisions, does not necessarily lead to a suppression of   pre-thermalization. However, the suppression will  be most effective in the presence of  SU$(N)$-breaking interactions provided the number of components $N \ge 4$ and the initial state contains a population imbalance that breaks the SU$(N)$ symmetry. We also find the conditions on the imbalance initial state that allow for a pre-thermal state to be stabilized for a certain time.  Our study highlights the important role played by the initial state in the pre-thermalization dynamics of multicomponent Fermi gases. It also demonstrates that alkaline-earth Fermi gases provide an interesting playground for the study and control of pre-thermalization. 
\end{abstract}
\maketitle

\section{Introduction}
Dilute Fermi gases of akaline-earth atoms (AEAs) like $^{173}$Yb or $^{87}$Sr exhibit a remarkable unitary symmetry at
ultracold temperatures ~\cite{Miguel2009,Gorshkov2010,Miguel2014}.
  The latter is ultimately a consequence of their closed-shell
(ground-state) electronic structure, $^1$S$_0$:  For AEAs, the atomic total angular momentum $\vec{F}$ in their ground state equals the nuclear spin $\vec{I}$. Due to the weakness of hyperfine interactions, $\vec{I}$ is essentially decoupled from the electronic degrees of freedom. Thus ultracold gases of AEAs can be viewed as spin-$I$ particles interacting with a pseudo-potential that is independent of their nuclear spin orientation, $I_z$ and therefore invariant under the larger unitary group   SU$(N=2I+1)$.  The accuracy of this SU$(N)$-symmetric description of interaction has been
 confirmed experimentally~\cite{Yamazaki2010,PhysRevA.84.043611_schreck}

  Recently, this property has attracted a great deal of interest in connection with the possibility of quantum emulation  of SU$(N)$-symmetric models of interest to  
condensed matter physicists~\cite{Affleck1988,Marston1989,PhysRevB.28.5255,Read1983,
Read1987,Read1989,Miguel2009,Gorshkov2010,PhysRevA.95.033619,Barbarino2015,Barbarino2016,PhysRevLett.112.043001, PhysRevA.93.051601,Hermele_magnetism,Mila_chiral,Mila_sun}. In recent years, many experiments along this direction have been carried out~\cite{Sugawa:2011aa_SUnexp,Taie:2012aa_SUnexp,PhysRevLett.121.225303_SUnexp,PhysRevX.6.021030_SUnexp,Zhang2014,Cappellini2014,Scazza2014} (See also~\cite{Miguel2014} for a recent review).  In addition, the realization of SU($N>2$) symmetric many-body systems is  relevant to the understanding of some aspects of the strong force that binds quarks into nucleons~\cite{OzawaBaym2010, Banerjee2013}. 
Indeed, as a quantum emulator of  $SU(N)$ interacting fermion gas for which $N$ can be as large as $10$, ultracold gases of AEAs can provide an ultracold realization of certain toy models of quantum chromodynamics (QCD)~\cite{Banerjee2013}. In this regard, it is interesting to explore any further connections between  ultracold gases of AEAs and quark-gluon physics. Indeed,  an idea that has recently emerged in the study of the quark-gluon plasma is the existence of  \emph{pre-thermalized} states~\cite{PhysRevLett.93.142002}.
The latter are  characterized by the rapid establishment of a kinetic temperature whilst, at the same time, the distribution of the eigenmodes of the system has not reached thermal equilibrium as described by the Fermi-Dirac or Bose-Einstein distributions. Pre-thermalized states have been extensively discussed in relation to the non-equilibrium dynamics of ultracold atomic gases~\cite{Moeckel2008,Moeckel2009,nessi_shorttime2014,
nopreth2d_2014,prethandth_2009_eckstein,Nessi_glass_2015,prethandth_silva_2013,turnable_integrablity_2014,nearintegrable_Lagen2016,Moeckel2010,spinprethermal2017,prethandth_mitra_2013,PhysRevLett.97.156403,preth_th_luttinger_2016,prethspinchain_Gong2013,GGEpreth_2011_kollar,Miguel2016,spin_prethandth_2015,nearintegrable_alba_2017,preth_spin_short_2018}. In earlier work,  it was also shown~\cite{nessi_shorttime2014} that pre-thermalization can be also linked to the existence of an integrable truncated version of the Hamiltonian that describes the short-time  dynamics~\cite{nessi_shorttime2014} and can be related~\cite{nessi_shorttime2014} to the non-thermal states occurring in integrable systems, which are described by the generalized Gibbs 
ensemble~\cite{rigol_gge,Miguel2016,
integrable_rigol_2007,isingchaing_2011,GGE_2015,1dboson_2014,PhysRevLett.97.156403,GGEpreth_2011_kollar,integrablebose_Andrei2016,1dbosepreth_kormos_2014,1dbose_kormos_2013}. However, for non-integrable systems,  the system will eventually  relax at long times to a thermal state described by a standard Gibbsian ensemble
~\cite{Rigol_ETH,rigol_breakintegrability_thermali_2010,rigol_breakintegrability_2010,rigol_thermalization_2009,rigol_thermalization2_2009,Biroli_thermalization_2010,thandpreth_2018,rigol_thermalization_2018,rigol_thermalization_NC_2018}. However, if the breaking of integrability is weak, the pre-themralized state  emerging at short times can be fairly long lived
~\cite{Moeckel2008,Moeckel2009,prethandth_2009_eckstein,eckstein_hubbard_2010,prethandth_silva_2013,turnable_integrablity_2014,nearintegrable_Lagen2016,Moeckel2010,spinprethermal2017,prethandth_mitra_2013,prethspinchain_Gong2013,GGEpreth_2011_kollar,spin_prethandth_2015,Nessi_glass_2015,boseprethramp2018,preth_spin1_2011,bose_prerh_cosme_2018,prethexp_Neye2016,bose_exp_Gring1318,bose_prethexp_2018,prethandth_2009_eckstein,kollarperturbation_2013,perturb_prethandth_2016,
nearintegrable_alba_2017}.

\begin{figure}[t]
\centering
\includegraphics[width=\columnwidth]{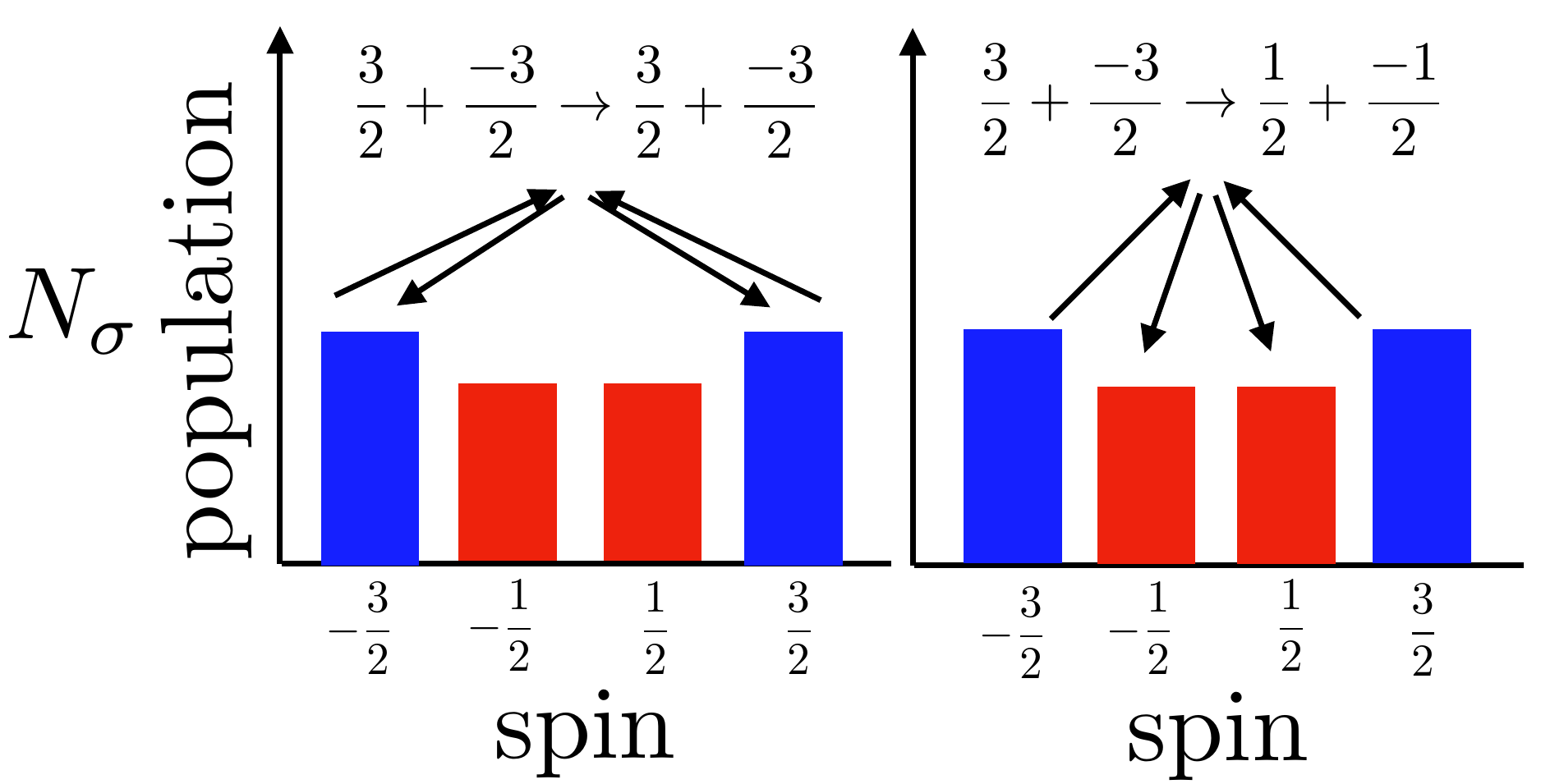}
\caption{Two possible scattering processes in an SU$(4)$ Fermi gas following a quench of the interaction that breaks the emergent SU$(4)$ symmetry, starting from an initial state with population imbalance. {\bf Right panel:} Spin-changing collision (as understood in this work) defines a two-atom collision that modifies the relative spin populations.
{\bf Left panel:} Other scattering processes that conserve total spin and do not change the relative spin-population.\label{SU(4)}}
\end{figure}

 Understanding the conditions under which a system exhibits pre-thermalization is an important ongoing research effort.  In this work, we investigate the control and suppression of the pre-thermalized behavior by the choice of the initial state and the symmetries of the post-quench Hamiltonian.  Mathematically, the emergent SU$(N = 2I +1)$ symmetry described above forbids spin-changing collisions that change the relative population of the different  spin components (cf. Fig.~\ref{SU(4)}). Thus, we shall consider   the short-time dynamics of  an  AEA Fermi gas following a sudden interaction quench in which the postquench Hamiltonian breaks the SU$(N)$ symmetry.  We analyze under which conditions this system exhibits pre-thermalization and, in particular, how the latter is affected by allowing for spin-changing collisions.  It is known~\cite{Moeckel2008,Moeckel2009,Moeckel2010, nessi_shorttime2014} that the
existence of pre-thermalized states in Fermi gases is related to the lack of phase space  for inelastic collisions to redistribute the excitations that are created following the quench. 
 Thus, thermalization is only possible after enough phase space has been created for inelastic collisions to efficiently take place~\cite{Moeckel2008,Moeckel2009,Moeckel2010}. The latter take place more efficiently as the strength of the quenched interaction is increased. This expectation has been confirmed numerically~\cite{eckstein_hubbard_2010,kollarperturbation_2013}. In this work we further analyze this issue and show that the mere existence of phase space allowing for inelastic collisions in the initial state is a necessary but \emph{not sufficient} condition for the suppression of pre-thermalization. Furthermore, the existence of matrix elements of the interaction allowing for inelastic collisions is not sufficient if it is not accompanied by the existence of phase space in the initial for the latter to take place. Nevertheless, as we show below, it is possible to find certain types of initial states for which the rate at which the inelastic collisions happens at intermediate times vanishes. 
   
The rest of this article  is organized as follows. The details of the model as well as the methods employed for the study of pre-thermalization  are discussed in Sec.~\ref{sec:model}. In Sec.~\ref{sec:appa}, the details of the calculations of the instantaneous momentum distribution are provided. In Sec.~\ref{sec:su2},
we describe the quench dynamics in the two-component gas. In Sec.\ref{sec:su4} we report the results for the four-component gas and discuss the effects of  spin-changing collisions and the spin-imbalance in the initial state. In Sec.~\ref{sec:conclu}, we provide the conclusions of this work and summarize the main results. Some details of the calculations are given in  Appendix~\ref{sec:appa}. Henceforth, we shall work in units where $\hbar = 1$.

\section{Model and Methods}\label{sec:model}
An interaction quench in an AEA ultracold gas can be described  by means of the following  Hamiltonian:
\begin{align}
H (t)&= H_0 + U_1(t),\\
H_0 &= K +  U_0,\label{ham}\\
K &= \sum_{\vec{p}\sigma}\epsilon_{p}c^{\dagger}_{\vec{p}\sigma}c_{\vec{p}\sigma},
\end{align}
where  $\epsilon_p = p^2/2m$ is the single-particle dispersion,  and $c_{\vec{p}\sigma},c_{\vec{p}\sigma}^{\dagger}$  are the annihilation/creation operators of fermions with momentum $p$ and spin $\sigma$ obeying $\{c_{\vec{p}\alpha},c_{\vec{k}\beta}^{\dagger}\}=\delta_{\alpha,\beta}\delta_{\vec{p},\vec{k}}$ (and anti-commuting otherwise); 
 $U_0$ is the initial (SU($N$)-symmetric~\cite{Miguel2009,Gorshkov2010,Miguel2014}) interaction, and 
$U_1(t) = \theta(t) U_1$ is the quenched interaction (see below), $\theta(t)$ being  Heaviside's step function, which
describes a sudden quench of the interaction term $U_1$. The generalization of our methods to other types of quenches 
has been presented in Ref.~\cite{Huang2019}, where special attention was paid to the proper definition
of the energy and the asymptotic behavior of the momentum distribution out of equilibrium. In the sudden
quench limit, the calcuation of dynamics of the total energy is less involved, see discussion at the end of this section.

 In general, for ultracold Fermi gases of spin-$F$ atoms, the
 generalization of Lee-Huang-Yang pseudo-potential that describes the two-particle collisions in the s-wave channel reads~\cite{Yip1998}:
\begin{align}
U_1&=  \sum_{J=0,2,..}^{2F-1} \frac{g_J}{2V} \sum_{M=-J}^J 
  \sum_{\alpha\beta\gamma\delta} \langle FF \alpha\beta | J M\rangle \langle JM |FF\gamma\delta\rangle 
\notag\\
&\qquad \times 
\sum_{\vec{pkqr}} 
 c_{\vec{p}\alpha}^{\dagger}c_{\vec{k}\beta}^{\dagger} c_{\vec{q}\gamma}
c_{\vec{r}\delta} \: \delta_{\vec{p+k,q+r}},\label{eq:U}
\end{align}
where $\langle \alpha\beta |FFmJ\rangle$ are  Clebsch-Gordan coefficients.  The couplings that parametrize the short-range interaction are  $g_{J}= 8 \pi   a^{J}_s/m$, where $a^{J}_{s}$ are the s-wave scattering lengths of the scattering channel with total spin $J$, $m$ is the atom mass and $V$ is the volume of the system. 
However, before the quench the AEAs in their ground state interact with an interaction for which all 
scattering lengths $a^J_s$ are identical with very high accuracy, which results in the emergent SU$(N)$ symmetry~\cite{Miguel2009,Gorshkov2010,Miguel2014} mentioned above. Thus, the initial interaction reads:
\begin{equation}
U_0= \frac{g_i}{2V}  \sum_{\vec{pkqr}}
 c_{\vec{p}\alpha}^{\dagger} c_{\vec{k}\beta}^{\dagger} c_{\vec{q}\beta}
c_{\vec{r}\alpha} \: \delta_{\vec{p+k,q+r}},
\end{equation}
where the coupling $g_i = 8\pi a_s/m$, $a_s$ being the scattering length.
 However,  suddenly turning on $U_1$ which contains different values of $g_J$ (i.e. $g_0 \neq g_2 \neq \ldots$) breaks the SU$(N)$ symmetry while respecting spin rotation symmetry. This means that in the two-particle scattering  events, the total (hyperfine) spin of the colliding particles is still conserved but the spins of the colliding particles can change. In the latter case, we speak of spin changing collisions (SCCs, see Fig.~\ref{SU(4)}, right panel).

 Unfortunately, for gases of AEAs the kind of interaction quench  envisaged above that breaks the emergent SU$(N)$ symmetry cannot be realized using magnetic Feshbach resonances because the latter are not accessible in the ground state due to their closed-shell electronic structure. However, by means of the so-called optical Feshbach resonances (OFR)~\cite{Blatt2011,Ciurylo2005,Enomoto2008,Yan2010,Yamazaki2010,PhysRevLett.77.2913_walraven,PhysRevLett.93.123001_hecker,PhysRevLett.105.050405_takahashi}  it is possible to (suddenly) enhance the values of the scattering lengths $a^J_s$.  To this end, a laser is used to couple a pair of colliding atoms with an excited bound state, which induces a Feshbach resonance and modifies the scattering lengths $a^J_s$. This method for enhancing the interaction violates  the emergent SU$(N)$ symmetry of the resulting interaction since the ground state is  coupled to an excited state that possess a hyperfine structure. As a result, it should possible to observe the dynamics of the initially SU$(N)$-symmetric gas that is subject to an SU$(N)$-symmetry breaking interaction quench.  The price to pay for the use of OFR is the introduction of inelastic losses, which result from the real  excitation to the excited bound state of a pair of AEAs. 
Inelastic losses will  provide an additional mechanism for the suppression of pre-thermalization. However, in our theoretical study, we shall neglect their effect in order to  understand the effects  of elastic interactions alone, which is a good approximation provided the gas is not driven too close to the resonant regime of the OFR. The latter condition is also consistent with the perturbative approach that we use below, which requires that the interaction couplings $g_J$ are  not too  large, and with the numerical observation that pre-thermalization is most observed in the weak to intermediate
coupling regime~\cite{eckstein_hubbard_2010,kollarperturbation_2013}.

   Following previous work~\cite{PhysRevLett.93.142002,Moeckel2008,nessi_shorttime2014,Miguel2016,Moeckel2009,Moeckel2010,perturb_prethandth_2016,PhysRevB.92.235135,GGEpreth_2011_kollar,perturb_werner2013,kollarperturbation_2013}, we shall study the evolution of the instantaneous momentum distribution and the total energy in order to identify the pre-thermalized regime, i.e. we compute:
\begin{align}
 n_{p\sigma}(t)&=\langle \Psi(t)| c^{\dagger}_{\vec{p}\sigma}c_{\vec{p}\sigma} | \Psi(t) \rangle,\\
E_{\mathrm{tot}}(t)&=\langle \Psi(t) | H(t)| \Psi(t) \rangle,
 \end{align}
where $|\Psi(t)\rangle$ is the solution of the time-dependent Schr\"odinger equation for the Hamiltonian $H(t)$ (cf. Eq.~\ref{ham}). In the following section, we compute the short-time dynamics of $n_{p\sigma}(t)$  using perturbation theory to the lowest non-trivial orders in the initial and quench  interactions. The second order results obtained below are  valid for times that fulfill the condition~\cite{Moeckel2008,Moeckel2009} $\epsilon_F t \lesssim (g^J_{\mathrm{max} } k_F^{3}/\epsilon_F)^{-3}  \simeq (k_F a^J_{\mathrm{max}})^{-3}$, where $g^J_{\mathrm{max}}$  ($a^J_{\mathrm{max}}$) is the largest coupling (scattering length) in the quenched interaction and $\epsilon_F$ ($k_F$) is the mean Fermi energy (momentum). 

 From the instantaneous momentum distribution, we obtain the total particle number of each spin-component, 
\begin{equation}
N_{\sigma}(t)=\sum_p n_{p\sigma}(t),
\end{equation}
as well the discontinuity of the momentum distribution at Fermi momentum:
\begin{equation}
 Z_{\sigma}(t)=\lim_{\delta\to 0^{+}} \left[ n_{p_{F\sigma}+\delta,\sigma}(t)-n_{p_{F\sigma}-\delta,\sigma}(t) \right].
\label{eq:zeq}
\end{equation}
Below, when discussing  pre-thermalization, we will  focus on $Z_{\sigma}(t)$ rather than on the full momentum distribution.

 For a sudden quench, the dynamics of total energy can be obtained by resorting to energy conservation. 
Consider the evolution of the total energy for times $t > 0$: The state of the system is described by  $|\Psi(t) \rangle = e^{-i (H_0 + U_1)t} |\Psi(0)\rangle$. Hence, 
\begin{align}
E_{\mathrm{tot}}(t >0) &= \langle \Psi(t)|  H(t)|\Psi(t) \rangle \\
&=  	\langle \Psi(0) | \left(H_0+U_1\right) |  \Psi(0)\rangle\\
 &= E_{0}  + \langle \Psi(0) | U_1 | \Psi(0)\rangle,\label{eq:nrg1}
\end{align}
where $E_{0} = \langle \Psi(0) | H_0 | \Psi(0)\rangle$.
In other words, the total energy is a constant for $t > 0$, and for any time $t$, it 
exhibits  rather simple evolution dynamics: 
\begin{equation}
E_{\mathrm{tot}}(t) = E_0 + \theta(t) \langle \Psi(0) | U_1 | \Psi(0)\rangle.
\end{equation}
The above result implies that the total energy immediately reaches its final (thermal) value after the quench. For a Dirac-delta
interaction (also called single-channel model), it is not possible to mathematically define the instantaneous kinetic energy. This is because  the instantaneous momentum distribution $n_{k\sigma}(t)$  behaves as $k^{-4}$ for $k \gg k_F$~\cite{Huang2019}, which renders the integral  $E_{\mathrm{kin}} = \sum_{\vec{k},\sigma} \epsilon_{k} n_{\vec{k}\sigma}(t)$ divergent. Thus, we shall define the pre-thermalized regime as a state in which  the total energy has reached its  (final) thermal value whilst the momentum distribution has not.  This means that
the existence of a pre-thermalized regime can entirely be inferred from the existence, for certain time following the quench, of a quasi-stationary, non-thermal momentum distribution. In Fermi systems, this is manifested by a plateau in the evolution of the instantaneous discontinuity at the Fermi momentum, $Z_{\sigma}(t)$ (cf. Eq.~\ref{eq:zeq}).
\section{Instantaneous momentum distribution}
 
\begin{figure}[t]
\center\includegraphics[width=0.9\columnwidth]{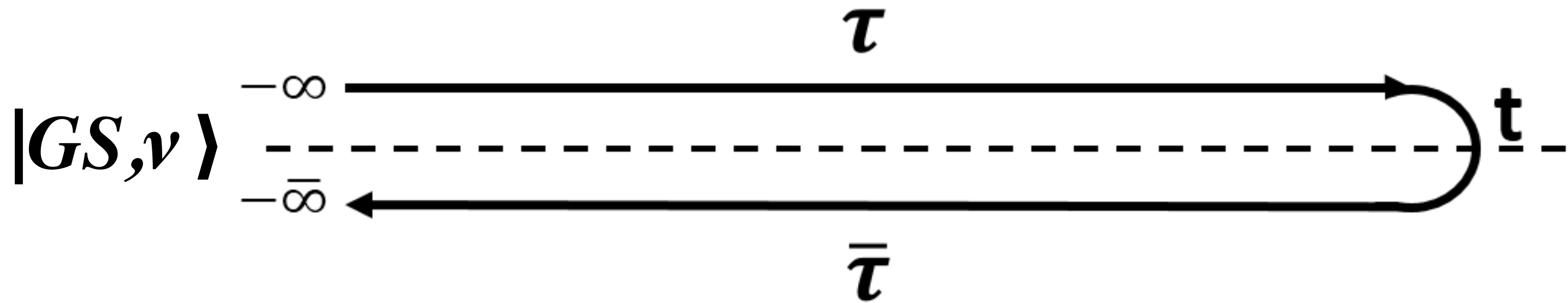}
\caption{\label{1} Closed-time contour $C$. Times $\tau$ and $\bar{\tau}$ lie on the time ordered and anti-time ordered branches, respectively. $\tau$ is earlier than $\bar{\tau}$ in contour ordering. The turning point $t$  is the time of at which observable in which we are interested is evaluated. $|GS,\nu\rangle$ is the initial state, which we take to be the ground state of non-interacting $N$-component Fermi gas characterized by a certain population ratio $\nu$ (see discussion below).}\label{contour}
\end{figure}

In this section, we describe how the time evolution of the instantaneous momentum distribution is obtained. Assuming that the interaction strength is weak, we shall compute the evolution  of a given  observable $O$ in a perturbative series in the total interaction, $V(t)= e^{iK t}\left[ U_0(t)+ U_1(t) \right] e^{-i K t}$, where we used that the initial interaction $U_0(t)=U_0 e^{-\eta|t|}$ is adiabatically switched on (off) at a rate $\eta\to 0^+$. Thus,
\begin{align}
O(t) &= \frac{\langle GS,\nu |\mathcal{T}[e^{-i \int_C dt\: V(t)} O(t)]|GS,\nu \rangle}{\langle GS,\nu |\mathcal{T}[e^{-i \int_C dt\: V(t)}]|GS,\nu \rangle}
\label{eq:opert}\\
&=\langle O\rangle\notag -i\int_C dt_1 \langle  \mathcal{T}\left[ V(t_1)O(t) \right] \rangle \notag \\
&\quad + \frac{(-i)^2}{2!}\int_C dt_1 dt_2 \, \langle  \mathcal{T} \left[ V(t_1)V(t_2)O(t) \right] \rangle_c\notag \\
&\quad + \cdots \label{eq:expan1}
\end{align}
where $\langle O \rangle = \langle GS, \nu | O  | GS, \nu \rangle$.  The times $t_1,t_2,\ldots$ all lie on the closed contour $C$ shown in Fig.~\ref{contour} and $\mathcal{T}$ is the time-ordering symbol on $C$. The state $ |GS, \nu \rangle$ denotes a non-interacting state, which is characterized by a particular ratio $\nu$ of the population of the different spin components (the definition  of $\nu$ depends on the number of components, see below).  Strictly speaking, the denominator of Eq.~\eqref{eq:opert} equals unity, but it is needed when expanding in powers of $V(t)$ in order to cancel disconnected terms resulting from the application of Wick's theorem to the above expression. 

 Previous work~\cite{Moeckel2008,nessi_shorttime2014,Miguel2016,Moeckel2009,Moeckel2010,perturb_prethandth_2016,PhysRevB.92.235135,GGEpreth_2011_kollar,perturb_werner2013,kollarperturbation_2013} has established  that pre-thermalization is accessible through perturbation theory.  In the problem of interest here,   the perturbation series for the observable $O$ can be organized as a double perturbative expansion in powers of $U_0$ and $U_1$, i.e.
\begin{align}
\langle O(t)\rangle
&=O^{(0,0)}+O^{(0,1)}(t)+ O^{(1,0)} \notag\\
&\quad +  O^{(1,1)} + O^{(0,2)}(t)+ O^{(2,0)}(t) + \ldots  
\end{align}
where $O^{(n,m)}(t)$ denotes the term  that is $n$-th order  in the initial interaction $U_0$ and $m$-th order in quenched interaction, $U_1$.   Next we set $O = c^{\dag}_{\vec{p}\sigma} c_{\vec{p}\sigma}$, and thus,
\begin{align}
n_{p\sigma}(t) &= n^{(0,0)}_{p\sigma}(t) + n^{(1,0)}_{p\sigma}(t)+ n^{(0,1)}_{p\sigma}(t) \notag \\
& + n^{(1,1)}_{p\sigma}(t) + n^{(2,0)}_{p\sigma}(t) + n^{(0,2)}_{p\sigma}(t) + \cdots  \label{eq:n}
\end{align}
where $n^{(0,0)}_{p\sigma}(t) = n^0_{p\sigma}$. In the closed time contour $C$, we 
can express the other $n^{(n,m)}_{p\sigma}(t)$ 
in terms  of  the self-energy and propagator matrices which, to second order in the total interaction, read:
\begin{align}
 n^{(1,0)}_{p\sigma}(t) &= \int_C  dt_1\:   \mathcal{G}_{p\sigma}(t,t_1) \Sigma^{(0,1)}_{p\sigma}(t_1)  \mathcal{G}_{p\sigma}(t_1,t),\notag\\
 n^{(0,2)}_{p\sigma}(t) &= \int_C  dt_1\:   \mathcal{G}_{p\sigma}(t,t_1) \Sigma^{(0,1)}_{p\sigma}(t_1) \mathcal{G}_{p\sigma}(t_1,t,)\notag\\
n^{(2,0)}_{p\sigma}(t) &=   \int_C  
dt_1 dt_2 \: \mathcal{G}_{p\sigma}(t,t_2)\:   \Sigma^{(2,0)}_{p\sigma}(t_2,t_1) \mathcal{G}_{p\sigma}(t_1,t),\notag\\
n^{(1,1)}_{p\sigma}(t) &=  \int_C  
dt_1   dt_2 \: \mathcal{G}_{p\sigma}(t,t_2)\:   
\Sigma^{(1,1)}_{p\sigma}(t_2,t_1) \mathcal{G}_{p\sigma}(t_1,t).
\end{align}
The propagator $\mathcal{G}_{p\sigma}(a,b)$ is defined in Eq.~\eqref{eq:g}, and $\Sigma^{(1)}_{p\sigma}(t_1)$, $\Sigma^{(2)}_{p\sigma}(t_2,t_1)$ can be computed using Feynman diagrams (see Appendix~\ref{sec:appa}).

 For the calculation of equal-time expectation values, we choose the time argument   of the  observable ($t$) to lie slightly before the turning point of the contour $C$,  which is on the time ordered ($\tau$) branch. In this 
case, the fermion propagators must be obtained from 
Eq.~\eqref{eq:g0}, which yields:
\begin{equation}
\mathcal{G}_{p\sigma}(t,b)%
=e^{-i\epsilon_{p}(t-b)}\begin{pmatrix} 1-n^0_{p\sigma}&-n^0_{p\sigma}\\0&0\end{pmatrix}\label{eq:Gg1},
\end{equation}
where the non-vanishing entries correspond to either $b$ lying before or after $t$ on the contour $C$. Similarly,
\begin{equation}
\mathcal{G}_{p\sigma}(a,t)=
e^{-i\epsilon_{p}(a-t)}\begin{pmatrix} -n^0_{p\sigma}&0\\1-n^0_{p\sigma}&0\end{pmatrix}.
\label{eq:Gg2}
\end{equation}
and the two non-zero entries correspond to $a$ lying before or after $t$ on the contour $C$. 
Combining Eq.~\eqref{eq:n}, the propagators, Eq.~\eqref{eq:Gg1} and Eq.~\eqref{eq:Gg2} and the second order corrections to the self-energy, Eqs.~\eqref{s1} to \eqref{s4}, we arrive at:
\begin{align}
n^{(2)}_{\vec{p}\sigma}(t) &=-\frac{2}
{V^2}\sum_{\alpha\beta\gamma}\sum_{\vec{pkqr}} Q^{\sigma\alpha\beta\gamma}_{\vec{pkqr}}
\int\limits_{-\infty}^{t} dt_1 \int\limits_{-\infty}^{t} dt_2\:   e^{i E_{pkqr}(t_1-t_2)} \notag \\
\times&\biggl[g^{(2)}_f(\sigma,\alpha;\beta,\gamma)\theta(t_1)\theta(t_2)+g_i^2\delta_{\sigma,\beta}\delta_{\alpha,\gamma}e^{-\eta(|t_1|+|t_2|)}\notag\\
&+2 g_i g_f^{(1)}(\sigma,\alpha;\sigma,\alpha)\delta_{\sigma\beta}\delta_{\alpha\gamma}\theta(t_1)e^{-\eta|t_2|}\biggr]
\label{eq:n2+0},\notag\\
&=-\frac{2}{V^2}\sum_{\vec{kqr}}\sum_{\alpha\beta\gamma} Q^{\sigma\alpha\beta\gamma}_{\vec{pkqr}}\biggl\{ \frac{g^2_i\delta_{\sigma\beta}\delta_{\alpha\gamma}}{E_{pkqr}^2}+\biggl[ g_{J}^{(2)}(\sigma,\alpha;\beta,\gamma)\notag\\
&\qquad+ g_i g_{J}^{(1)}(\sigma,\alpha;\sigma,\alpha)\delta_{\sigma\beta}\delta_{\alpha\gamma}\biggr] F(E_{pkqr},t)\biggr\},
\end{align}
where we have defined $E_{pkqr}=\epsilon_p+\epsilon_k-\epsilon_q-\epsilon_r$ and introduced $F(E,t)$ to denote the result  of the integration over $t_1$ and $t_2$:
\begin{equation}
F(E,t)=\frac{4\sin^2\left(Et/2\right)}{E^2}.\label{eq:FS2}
\end{equation}
In Eq.~\eqref{eq:n2+0} $Q^{\sigma\alpha\beta\gamma}_{pkqr}$ corresponds to the following expression:
\begin{align}
Q^{\sigma\alpha\beta\gamma}_{\vec{pkqr}} &=\delta_{\vec{p+k,q+r}}
 \biggl[n^0_{p\sigma}n^0_{k\alpha}(1-n^0_{q\beta})(1-n^0_{r\gamma})\notag \\ 
& \quad   -(1-n^0_{p\sigma})(1-n^0_{k\alpha})n_{q\beta}^0n_{r\gamma}^0\biggr].
\end{align}

\section{Two-component Fermi gas}\label{sec:su2}
%
%
\begin{figure}[b]
\centering
\includegraphics[width=0.4\columnwidth]{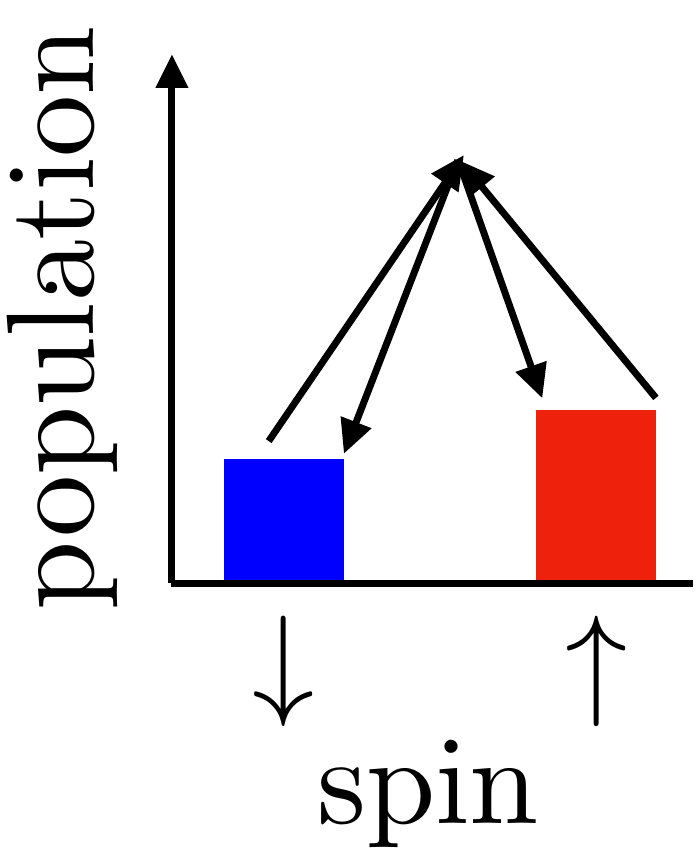}
\caption{Schematic representation of the only scattering process allowed in a two-component gas with Dirac-delta  interactions. Notice that $(\uparrow+\downarrow)\to(\uparrow+\downarrow)$, thus preserving the spin-populations for each spin component.}\label{SU(2)}
\end{figure} 
\begin{figure}[t]
\centering
\includegraphics[width=\columnwidth]{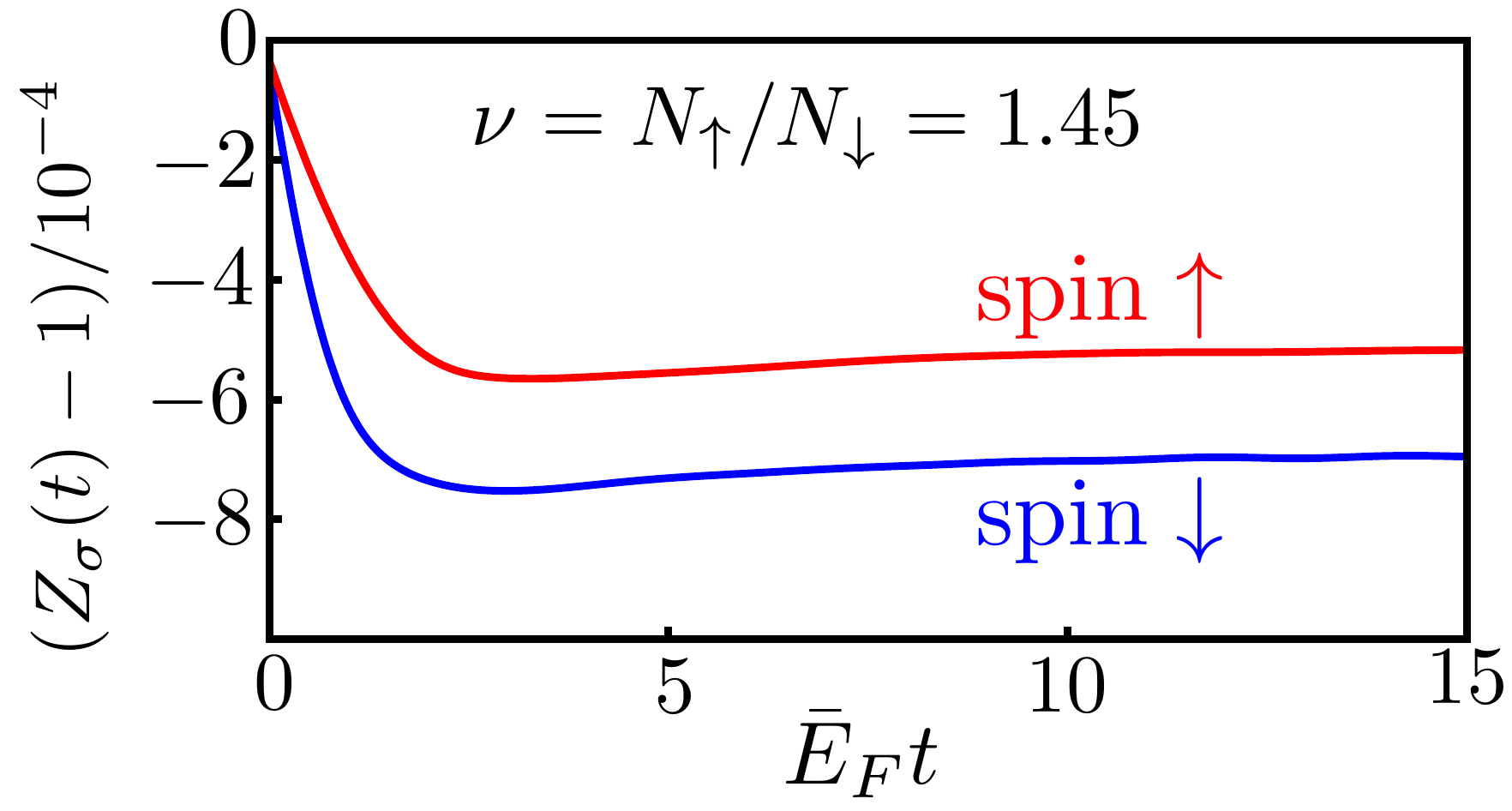}
\caption{(color online)
 Dynamics of discontinuity of the momentum distribution at Fermi momentum $Z_{\sigma}(t)$ after a sudden interaction quench in a two-component Fermi gas.  After a short transient, a plateau is observed indicating the existence of a pre-thermalized regime.
Time is measured  in units of the inverse of mean Fermi energy, $\bar{E}_F=(\epsilon_F^{\uparrow}+\epsilon_F^{\downarrow})/2=\bar{k}^2_F/2m$. The interaction strength is 
$\bar{k}_F a^0_s=0.0158$ for the quenched interaction and we take $\bar{k}_F a_s=0.0097$ for the initial interaction.}\label{SU(2)result}\label{fig:n2}
\end{figure} 
 As a warm up, in this section we consider the two component gas in order to show that breaking the SU$(N= 2)$ symmetry of the initial state by introducing 
an initial population imbalance does not suppress 
pre-thermalization. 

 Let us first recall that for a contact interaction in a two-component system, the pseudo-potential in Eq.~\eqref{eq:U} is parametrized by a \emph{single} coupling  constant only, which is determined by the s-wave scattering length for atoms colliding with total $J = 0$, $a_s$. Therefore,  it is not possible to break the SU$(N=2)$ symmetry of the interaction. Indeed, as shown in Fig.~\ref{SU(2)}, 
the only scattering process in a two-component gas is 
of the type $(\uparrow+\downarrow)\to(\uparrow+\downarrow)$. This kind of scattering process preserves the  populations of each spin component.  Thus, even if the initial state contains an population imbalance  (i.e. for
$\nu=N_{\uparrow}/N_{\downarrow}\neq 1$), the time evolution following the quench cannot alter the ratio $\nu$. This conservation law  alone  protects the existence of the pre-thermalized regime.

 In order to show that pre-thermalization is preserved,  we  have evaluated  explicitly the time evolution of the discontinuity of the instantaneous momentum distribution at Fermi momenta, i.e. $Z_{\sigma}(t)$, see Fig.~\ref{SU(2)result}.  Notice that, after a short transient,  $Z_{\sigma}(t)$ displays a plateau for both spin components. This is a behavior characteristic of the pre-thermalized regime indicating that the momentum distribution is non-thermal, similarly to what was found in previous studies using different models and spin-unpolarized initial states~\cite{PhysRevLett.93.142002,Moeckel2008,nessi_shorttime2014,kollarperturbation_2013}. The initial population imbalance is reflected in the pre-thermal value of  $Z_{\uparrow}(t)$ being different  from that of $Z_{\downarrow}(t)$. 
Indeed, we have checked that this  result also applies to $N > 2$-component Fermi gas. Thus, we conclude that, if the post-quench Hamiltonian retains the SU$(N)$ symmetry, the system shows pre-thermalization independently of the existence of population imbalance in the initial state (see discussion in the following section).

 %


\begin{figure}[h]
\centering
\includegraphics[width=\columnwidth]{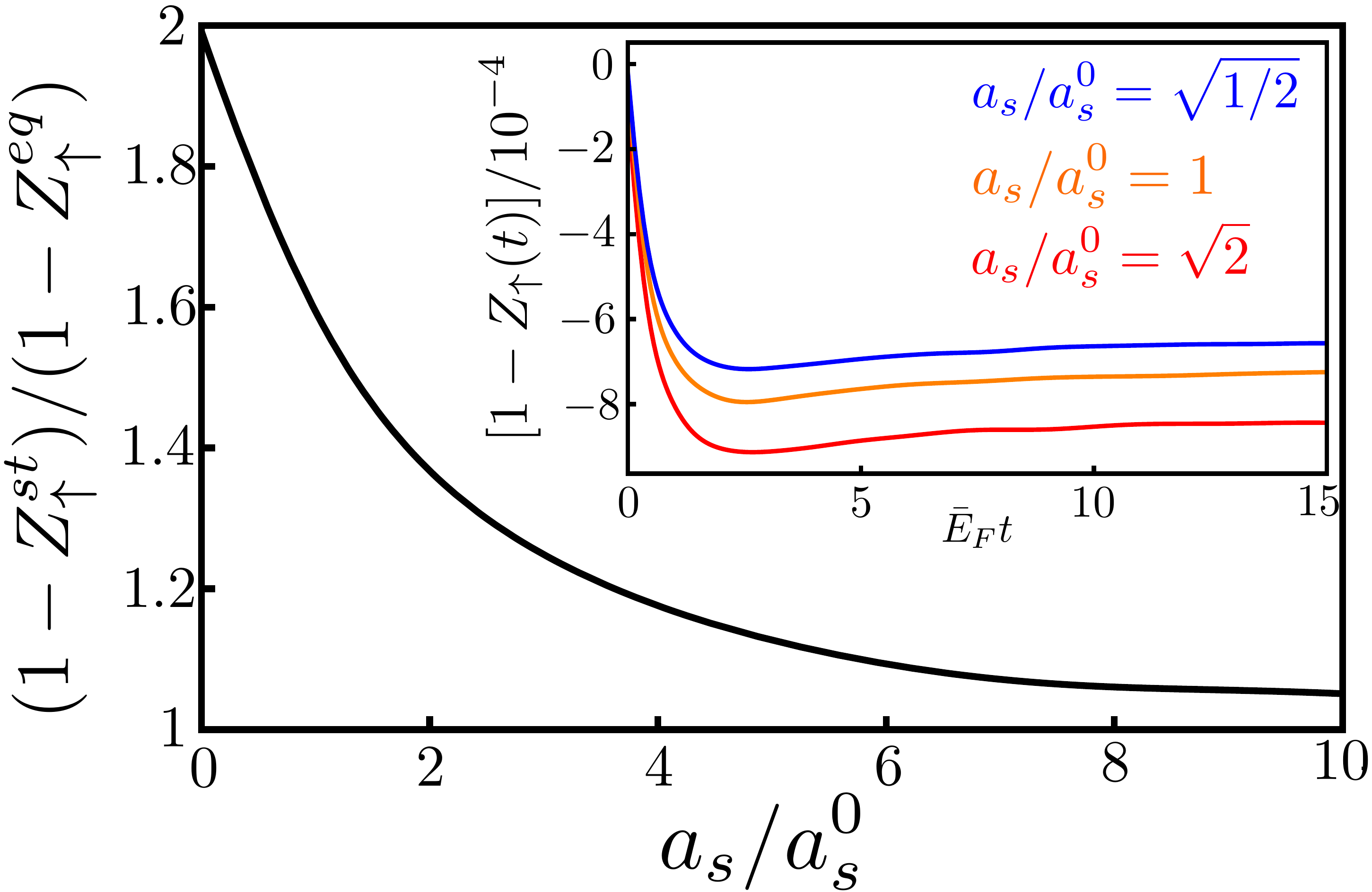}
\caption{Ratio of the stationary  and  equilibrium values of the discontinuity at the Fermi momentum of the momentum distribution, $Z_{\sigma}$, for the spin $\sigma=\uparrow$ in  a two-component Fermi gas following a sudden interaction quench where $a^0_s$ is the scattering length of the quenched interaction, and $a^0$ is the scattering length of the initial interaction. The post-quench total scattering length is $a_s+a_s^0$. As discussed in the main text, the ratio $(1-Z^{st}_{\uparrow})/(1-Z^{eq}_{\uparrow})$  ranges between $1$ and $2$ and depends on the ratio of the initial to the final interaction strengths. The inset shows the evolution of discontinuity at Fermi momentum for different values of the ratio $a_s/a^0_s$. The interaction strength for the quenched interaction is $\bar{k}_F a^0_s=0.0158$. Time is measured in units of the inverse of the mean Fermi energy $\bar{E}_F=(\epsilon_F^{\uparrow}+\epsilon_F^{\downarrow})/2=\bar{k}^2_F/2m$.}\label{fig:c}
\end{figure} 

 Next we consider the effects of the initial interaction. Although the latter is  SU$(N=2)$ symmetric and therefore will not suppress pre-thermalization, for the sake of experimental interest, it  is worth analyzing its  quantitative effect on  the dynamics of the instantaneous momentum distribution.  As a function of the ratio $a_s/a^0_s$ Fig.~\ref{fig:c} shows the ratio of the pre-thermalized value of $Z_{\uparrow}(t)$ to its value in the ground state of $H_0 + U_1 = K + U_0 + U_1$ (see Appendix~\ref{sec:appa} for the details of the calculation). Recall that the (initial) interaction strength is $g_i=8\pi a_s/m$ and $a^0_s$ is the scattering length characterizing the quench interaction strength $g_0=8\pi a^0_s/m$. Thus, the final interaction  is proportional to  $8\pi (a_s+a^0_s)/m$. The inset shows the full time dependence for a few values of  the ratio $a_s/a^0_s$. The results shown in Fig.~\ref{fig:c} can be summarized by the following relation:
\begin{align}
\frac{1-Z^{\mathrm{st}}_{\sigma}}{1-Z_{\sigma}^{eq}}=C\left(\frac{a_s}{a^0_s}\right),
\end{align}
where the crossover function $C(x)$ takes the following limiting forms: $C(x\gg1)=1$ and $C(x\ll1)=2$, while interpolating in between for intermediate values of $x = a_s/a^0_s$. The $x\to 0$ limit, which corresponds to the non-interacting initial state, has been obtained in previous work~\cite{nessi_shorttime2014,Moeckel2008,Moeckel2009}.
\begin{figure*}[t]
\centering
\includegraphics[width=\textwidth]{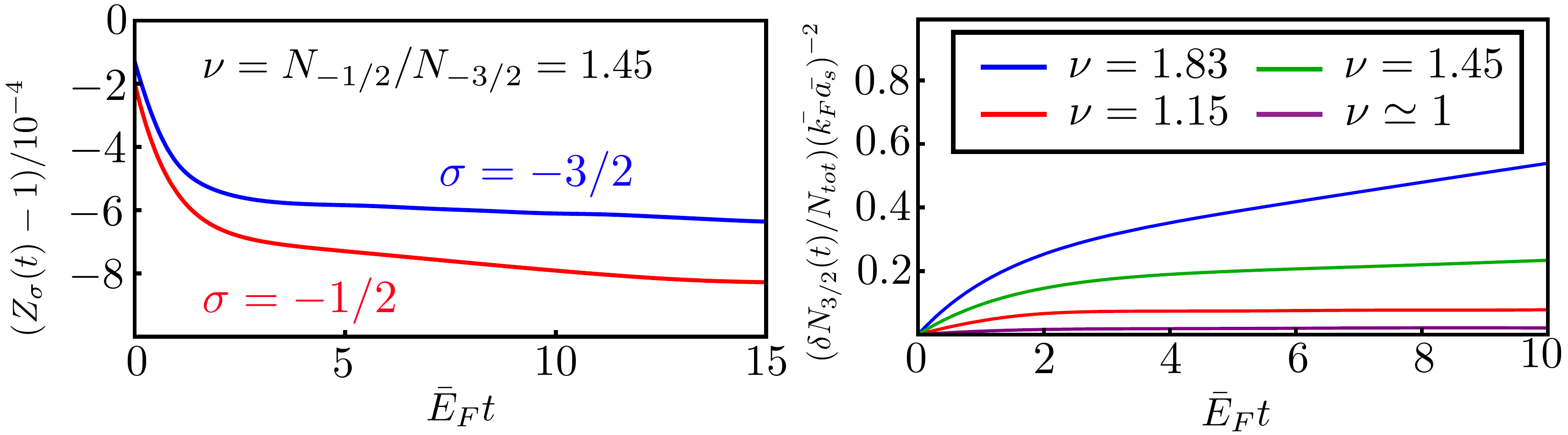}
\caption{(color online) {\bf Left panel: } Dynamics of the discontinuity at Fermi momentum of the momentum distribution for a $N=4$ Fermi gas following a sudden quench of the interaction that breaks the SU$(4)$ symmetry. The population ratio in the initial state $\nu=N_{\pm 1/2}/N_{\pm 3/2}=1.45$. {\bf Right panel: } Evolution of the population ratio $\delta N_{\sigma}/N_{tot} (\bar{k_F}\bar{a}_s)^{-2}$ for $\sigma= 3/2$, and different initial conditions with symmetry ($N_{1/2}=N_{-1/2}$, $N_{3/2}=N_{-3/2}$).  Time is measured in units of the inverse of the mean Fermi energy $\bar{E}_F=(\epsilon_F^{3/2}+\epsilon_F^{1/2})/2=\bar{k}^2_F/2m$. The interaction strength determined by the parameters $\bar{k}_F a^0_s=0.0158$ and $\bar{k}_F a^2_s=0.0032$, for the quenched interaction, and $\bar{k}_F a_s=0.0097$ for the initial interaction, respectively.} 
\label{fig:su4result}
\end{figure*}

\section{Suppression of pre-thermalization}\label{sec:su4}
%
%

 The dynamics of the two-component gas described above illustrates that introducing  a population imbalance in the initial state is not  a sufficient condition to suppress pre-thermalization. In this section, we  show that the situation changes dramatically when, besides the initial population imbalance, we allow for spin-changing collisions (SCCs) that result from the breaking of the SU$(N)$ symmetry  of the quenched interaction. 
For a contact interaction that conserves the total angular momentum, the minimum value of $N$ allowing for SCCs  is $N = 4$.  Fig.~\ref{SU(4)}  schematically shows  the two possible types of scattering processes. They are classified into two types: those preserving the relative spin populations (left panel) and the SCCs (right panel). 

 Next, we consider a sudden interaction quench in an initially interacting (with scattering length $a_s$) four-component Fermi gas. We assume the initial state contains an imbalance in the population of the different spin components as shown in Fig.~\ref{SU(4)}. This  population imbalance can be parametrized by the ratio $\nu = N_{\pm1/2}/N_{\pm3/2}$, which  also determines the non-interacting state $|GS,\nu\rangle$  used in the perturbative treatment outlined in Sec.~\ref{sec:appa}.  On the left panel of Fig.~\ref{fig:su4result} we have plotted the time evolution of  $Z_{\sigma}(t)$  for  $\nu=1.45$. In this case, unlike the plateau observed for the \emph{imbalanced} two-component gas, we find a slow decay of $Z_{\sigma}(t)$  as a function of time, which is indicative of the absence of pre-thermalization.  In order to shed further light into the behavior of the system, it is useful to consider the dynamics of the ratio $\delta N_{\pm 3/2}(t)/N_{\mathrm{tot}}$, where $\delta N_{\sigma}(t)$ measures the deviation from its initial value of the population for spin component $\sigma$.   This ratio is shown in Fig.~\ref{fig:su4result}, which illustrates how the  SCCs alter the relative populations by decreasing the population of the majority components with $\sigma = \pm 1/2$ and increasing that of the minority component (note that $\delta N_{\pm 1/2} (t) =-\delta N_{\pm 3/2}(t)$ because the total number is conserved). This change in relative population is the driving force behind the change of the kinetic energy for the  components with $\sigma = \pm 3/2$ and $\sigma = \pm 1/2$.  
 
   Analytically (see Appendix~\ref{app:b}), it can be shown  that, after a short transient, the rate of population change is given
by Fermi's Golden rule. This result can be obtained by formally taking the limit $t\to +\infty$ of the second order expressions for $\delta N_{\sigma}(t)$, i.e.
\begin{align}\label{eq:N}
&\lim_{t\rightarrow\infty} \frac{\delta N_{\sigma}(t) }{t}\propto \frac{1}{V^2}\sum_{\vec{pkqr}} 
  \sum_{\alpha\beta\gamma} g^{(2)}_{f} (\sigma,\alpha;\beta,\gamma)\left[n^0_{p\sigma}n^0_{k\alpha}\right.\notag\\
&\left.\times (1-n^0_{q\beta})(1-n^0_{r\gamma})-(1-n^0_{p\sigma})(1-n^0_{k\alpha})n^0_{q\beta}n^0_{r\gamma}\right]\notag\\
&\times \delta(\epsilon_p+\epsilon_k-\epsilon_q-\epsilon_r) \delta_{\vec{p+k,q+r}},
\end{align}
where $g^{(2)}_{f}(\sigma,\alpha;\beta,\gamma)$ is the (square of the)  matrix element for SCCs. 

Note that the rate of population change is independent of the initial SU$(N)$-symmetric interaction. The slope is proportional to the phase space volume available for SCCs multiplied by the matrix element mediating the transitions. Thus, this rate can be used as a measure of how severe the suppression of pre-thermalization is. As shown in Fig.~\ref{fig:su4result},  the rate of change of $\delta N_{\sigma}(t)$ (that is, its slope in Fig.~\ref{fig:su4result}) becomes larger as the initial population imbalance increases, as expected from the corresponding enhancement in the available phase space. 
Since the SCCs induce inelastic scattering processes, they 
provide a decoherence mechanism in the short time dynamics after the quench. Thus, for an initial state with $\nu\neq1$ in the presence of SCCs, neither the relative population  nor  the momentum distribution for each spin component become stationary after a short-time transient, indicating that  the  intermediate time behavior cannot be described  as pre-thermal. Conversely, the pre-thermalized regime is robust when inelastic scattering processes are (Pauli-)blocked at short time, which is the case of an initial state with no spin imbalance (i.e. $\nu = 1$), or
when the SCCs are absent because the quenched interaction is SU$(N)$ symmetric.

\section{Control of pre-thermalization}

We have seen in the previous sections that pre-thermalization can take place provided the symmetries of the
post-quench Hamiltonian or  the initial state are properly broken. Here we demonstrate the possibility of 
controlling the existence of pre-thermalization by carefully choosing the initial state. 

 As mentioned above,  Eq.~\eqref{eq:N} shows that the rate of change of the relative populations vanishes when the population of the different spin species is the same and the initial state becomes SU$(N=4)$ symmetric. However, this is not 
the only type of initial condition for which the rate of change of relative populations vanishes. Indeed, it is possible to find
other types of  initial states for which the rate of change of the relative population, Eq.~\eqref{eq:N} vanishes.  To see this, let us define $E_{p\alpha} = \epsilon_{p} - \epsilon^{\alpha}_F$, where $\epsilon^{\alpha}_F$ is the 
Fermi energy for the component $\alpha$.  Thus, the Dirac delta function, ensuring energy conservation in
Eq.~\eqref{eq:N} becomes:
\begin{equation}
\delta(\epsilon_p+\epsilon_k-\epsilon_q-\epsilon_r) = \delta(E_{p\sigma}+E_{k\alpha}-E_{q\beta}-E_{r\gamma} + \Delta_F)
\end{equation}
where $\Delta_F = \epsilon^{\sigma}_F+\epsilon^{\alpha}_F - \epsilon^{\beta}_F - \epsilon^{\gamma}_F$. In addition, we notice that $\delta N(t)$ receives contributions only from the SCCs, which for $N  = 4$ means that $\sigma = -\alpha = \tfrac{3}{2}$ and $\beta = -\gamma = \tfrac{1}{2}$. Thus, 
\begin{equation}
\Delta_F =  \epsilon_F^{+3/2}+\epsilon_F^{-3/2}-\epsilon_F^{+1/2}-\epsilon_F^{-1/2}.\label{eq:criteria}
\end{equation}
Next we shall argue that initial states satisfying $\Delta_F = 0$, will exhibit pre-thermal behavior (when initial interactions
are present, the initial state must be adiabatically connected to a non-interacting state that satisfies the condition $\Delta_F   =0$ since the initial SU($N)$-symmetric interaction does not induce SCCs). In order to establish this result, we first
notice that the expression in Eq.~\eqref{eq:N} contains two terms. In the first  one, the occupation factors 
$n^0_{p\sigma}n^0_{k\alpha} (1-n^0_{q\beta})(1-n^0_{r\gamma})$ require, at $T = 0$, that is, for a pure state, that 
$E_{p\sigma},E_{k\alpha} \ge 0$ and $E_{q\beta},E_{r\gamma} \leq 0$. At the same time, the energy conservation
for $\Delta_F  =0$ requires that $E_{p\sigma} + E_{k\alpha} = E_{q\beta} + E_{r\gamma}$, which can only be
satisfied if $E_{p\sigma}  = E_{k\alpha} = E_{q\beta}  =  E_{r\gamma} = 0$.  The manifold of points satisfying the
previous  condition in the nine-dimensional space span by the vectors $\vec{p},\vec{k},\vec{q}$ 
($\vec{r} = \vec{p}+\vec{k}  -\vec{r}$ is fixed by  momentum conservation) is a set of zero measure
and does not contribute to the integrals over momentum in Eq.~\eqref{eq:N}. An entirely identical
conclusion is reached for the second term in Eq.~\eqref{eq:N}. Physically, the condition that $\Delta_F = 0$
amounts to having  zero phase space for the SCCs to occur, even if this is not obvious from the
fact that the initial state contains a population imbalance.  This argument can be easily generalized to the case
 where $N > 4$.

\begin{figure*}[t]
\includegraphics[width=\textwidth]{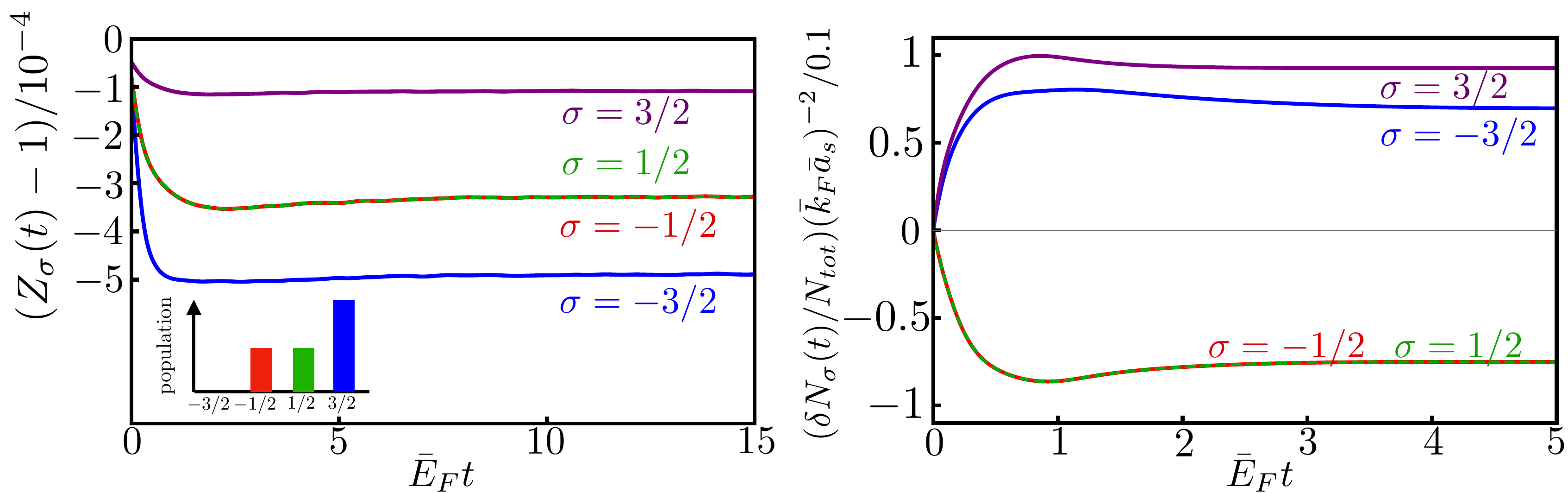}
\caption{(color online) {\bf Left panel:} Discontinuity at the Fermi momentum starting from an imbalanced initial state. The plateau indicates the emergence of pre-thermalization in the presence of SCCs. The insertion shows the initial condition schematically where the imbalanced initial state is 
 $\epsilon_F^{-3/2}=0$ and $2\epsilon_F^{\pm1/2}=\epsilon_F^{3/2}$. 
 {\bf Right panel:} Time-evolution of the relative population. After an initial exchange of particles, the spin populations reach stationary values. Time is measured in units of the inverse of the mean Fermi energy $\bar{E}_F=\frac{1}{4}\sum_{\sigma}\epsilon_F^{\sigma}=\bar{k}_F^2/2m$. The interaction strengths satisfy $\bar{k}_F a^0_s=0.0158$ and $\bar{k}_F a^2_s=0.0032$ (for the quenched interaction) and $\bar{k}_F a_s=0.0097$ (for the initial interaction).}\label{fig:cZ}
\end{figure*}

 In order to explicitly show how pre-thermal behavior emerges when $\Delta_F = 0$, we have computed the evolution of
$Z_{\sigma}(t)$ and $\delta N_{\sigma(t)}$  by numerically evaluating the corresponding expressions for an initial
state satisfying the condition that $\Delta_F = 0$. 
The results are shown in Fig.~\ref{fig:cZ}. It is worth comparing the results on the right panel of  Fig.~\ref{fig:cZ}, 
with those shown in Fig.~\ref{fig:cZ} (left panel). It can be seen that, after a short transient, for the initial
state satisfying the condition $\Delta_F = 0$, the curves for $\delta N_{\sigma(t)}$  flatten out. Concurrently, 
$Z_{\sigma}(t)$ also reaches the characteristic pre-thermal plateau. However, it is worth noticing that unlike the case of initial states with SU$(N)$ symmetry which trivially satisfy $\Delta_F = 0$, there is a change in population during the short-time transient because there are SCCs that do not satisfy energy conservation. The energy conservation is only enforced for $t$ sufficiently large. When this happens, the system enters the pre-thermalized regime. 
As a caveat, it is important to notice that if change of population happens during the short-time transient, then 
condition $\Delta_F = 0$ will not ensure that the system reaches the pre-thermalized regime. 
Thus, we must require that $\delta N_{\sigma}(t)/N_{\sigma}\ll1$, which is the case in the regime where 
quenched interaction can be treated perturbatively. 

\section{Conclusions}\label{sec:conclu}
 In conclusion, we have studied the pre-thermalization dynamics of an isolated multi-component Fermi gas of ultracold  atoms following a sudden interaction quench. This type of nonequilibrium dynamics can be experimentally studied using  alkaline earth atoms, whose interaction can be tuned using optical Feshbach resonances~~\cite{Blatt2011,Ciurylo2005,Enomoto2008,Yan2010,Yamazaki2010}, e.g. in $^{173}$Yb for $N\leq 6$ and $^{87}$Sr for $N\leq 10$. 
 
   We have shown that the short-time dynamics of this system is affected by both the presence of a population imbalance in the initial state and the breaking of the emergent SU($N$) symmetry of the contact interactions by allowing spin-changing collisions.  Both elements are necessary for the suppression of pre-thermalization, as illustrated by 
the behavior of  two-component Fermi gas with  initial spin imbalance, which displays a robust pre-thermalizated regime at short to intermediate times. On the other hand, for a generic SU($N\ge 4$) symmetry-breaking (i.e. imbalanced) initial state we do not observe a pre-thermal  regime after a quenching an SU$(N)$-symmetry-breaking interacting. This is because, generically, the  population imbalance in the initial state provides phase space for inelastic spin-changing collisions and introduces a decoherence mechanism and suppress pre-thermalization. 

Nevertheless, we have shown that there is a 
class of imbalanced initial states which allows for the emergence of pre-thermal behavior. This opens the possibility of
using multi-component gases to study the suppression and control of this nonequilibrium state. 
In addition, the findings reported in this work should allow  for the possibility to experimentally observe 
the effect of SCCs by studying the dynamics of the quantum gas following an interaction quench.

\acknowledgments

MAC and CHH have been supported by the Ministry of Science and Technology (Taiwan) under contract numbers NSC 102- 2112-M-007-024-MY5 and   107-2112-M-007-021-MY5. MAC also acknowledges the support of the National Center for Theoretical Sciences  (NCTS) of Taiwan. YT and YT acknowledge the supports by the Grant-in-Aid for Scientific Research of the Ministry of Education, Culture
Sports, Science, and Technology / Japan Society for the
Promotion of Science (MEXT/JSPS KAKENHI) Nos.
25220711 and 17H06138, 18H05405, and 18H05228; the
Impulsing Paradigm Change through Disruptive Tech-
nologies (ImPACT) program; Japan Science and Tech-
nology Agency CREST (No. JPMJCR1673), and MEXT Quantum Leap Flagship Program (MEXT Q-LEAP)(JPMXS0118069021).

\appendix
\section{Technical details of the calculations}\label{sec:appa}
\begin{figure}[b]
\center
\includegraphics[width=\columnwidth]{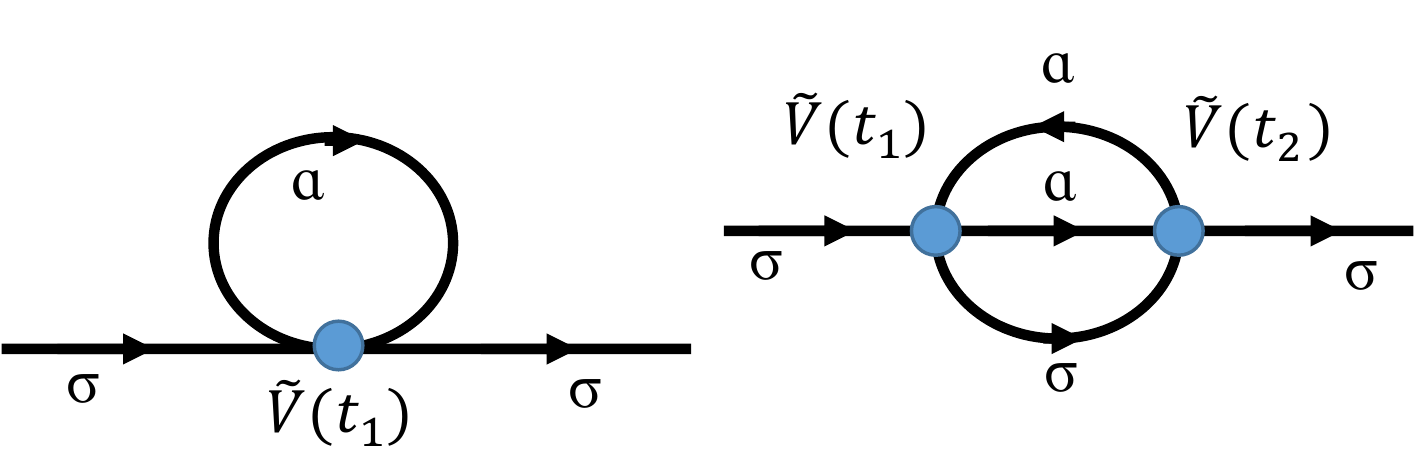}
\caption{First and second order diagram for momentum distribution. $V(t) = U_0(t) + U_1(t)$ is the sum of the total (initial plus quenched) interaction.}
\label{fig:feyn1}
\end{figure}

The fully connected contributions (denoted by $\langle \ldots \rangle_c$ in Eq.~\ref{eq:expan1}) resulting from the application of Wick's theorem can be represented in terms of Feynamn graphs (cf Fig.~\ref{fig:feyn1}). Notice that there are four possible choices for thhe time arguments for the fermion propagator:
\begin{equation}
i G_0(t_1,t_2;p\sigma)=  \langle \mathcal{T} \left[ c_{\vec{p}\sigma}(t_1)c_{\vec{p}\sigma}^{\dagger}(t_2)\right]\rangle, \label{eq:g0}
\end{equation}
where $t_1$ and $t_2$ can be either in the $\tau$ or $\bar{\tau}$ branches of $C$.  The free fermion propagator can be written in matrix form as follows:
\begin{align}
&\mathcal{G}_{p\sigma}(a,b)=
\begin{pmatrix}
iG^{T}_{p\sigma}(a,b)&iG^{<}_{p\sigma}(a,\bar{b})\\
iG^{>}_{p\sigma}(\bar{a},b)&iG^{\tilde{T}}_{p\sigma}(\bar{a},\bar{b})
\end{pmatrix}.\label{eq:g}
\end{align}
Using  $c_{\vec{p}\sigma}(t)=c_{\vec{p}\sigma}e^{-i\epsilon_{p}t}$, the entries of above matrix can be evaluated to yield:

\begin{align}
i G_{p\sigma}^{<}(t_1,\bar{t}_2)&= -n_{p\sigma}^0e^{i\epsilon_p(t_2-t_1)},\label{eq:gg1}\\
i G_{p\sigma}^{>}(\bar{t}_1,t_2)&=(1- n^0_{p\sigma})e^{i\epsilon_p(t_2-t_1)},\label{eq:gg2}\\
i G^{T}_{p\sigma}(t_1,t_2)&=
\theta(t_1-t_2) iG_{p\sigma}^{>}(\bar{t}_1,t_2)\notag\\
&+\theta(t_2-t_1)iG^{<}_{p\sigma}(t_1,\bar{t}_2;),\label{eq:g3}\\
i G^{\tilde{T}}_{p\sigma}(\bar{t}_1,\bar{t}_2)&=
\theta(t_2-t_1) iG^{>}_{p\sigma}(\bar{t}_1,t_2)\notag\\ 
&\quad +\theta(t_1-t_2)iG^{<}_{p\sigma}(t_1,\bar{t}_2).\label{eq:g4}
\end{align}

 For the calculation of equal-time expectation values, we choose the time argument  of the  operator ($t$) to lie slightly before the turning point of the contour $C$,  which is on the time ordered ($\tau$) branch. In this 
case, the fermion propagators must be obtained from 
Eq.~\eqref{eq:g0}, which yields:
\begin{equation}
\mathcal{G}_{p\sigma}(t,b)%
=e^{-i\epsilon_{p}(t-b)}\begin{pmatrix} 1-n^0_{p\sigma}&-n^0_{p\sigma}\\0&0\end{pmatrix}\label{eq:G1},
\end{equation}
where the non-vanishing entries correspond to either $b$ lying before or after $t$ on the contour $C$. Similarly,
\begin{equation}
\mathcal{G}_{p\sigma}(a,t)=
e^{-i\epsilon_{p}(a-t)}\begin{pmatrix} -n^0_{p\sigma}&0\\1-n^0_{p\sigma}&0\end{pmatrix}.
\label{eq:G2}
\end{equation}
and the two non-zero entries correspond to $a$ lying before or after $t$ on the contour $C$. 

The self-energy can be calculated from the diagrams shown in Fig.~\ref{fig:feyn1} and the propagators, Eqs.~\eqref{eq:gg1} to \eqref{eq:g4}. Thus, to first order in $V(t)$, we obtain:
\begin{align}
&\Sigma^{(0,1)}_{\sigma}(t_1)=\frac{\theta(t_1)}{V}
\sum_{\alpha}g_f^{(1)}(\sigma,\alpha;\sigma,\alpha)
\sum_{\vec{k}}n^0_{k\alpha}\label{eq:s01},
\end{align}
\begin{align}
&\Sigma^{(1,0)}_{\sigma}(t_1)=\frac{g_i}{V}
\sum_{\alpha\neq\sigma}
\sum_{\vec{k}}n^0_{k\alpha}e^{-\eta|t_1|}\label{eq:s10},
\end{align}
where we have introduced:
\begin{equation}
g_f^{(1)}(\sigma,\alpha;\sigma,\alpha)=\sum_{J=0,2..}\sum_{M}g_J  \langle J M|FF\sigma\alpha\rangle\langle FF \sigma\alpha|JM\rangle.\label{eq:g1}
\end{equation}
Combining the expression in matrix form, Eq.~\eqref{eq:n}, the propagators, Eq.~\eqref{eq:G1} and Eq.~\eqref{eq:G2}, and the self-energy for the first order correction, Eqs.~\eqref{eq:s01} and \eqref{eq:s10},
we obtain that first order correction to the instantaneous momentum distribution vanishes, i.e. $n_{p\sigma}^{(1)}=0$.

 At second order in the quenched interaction, we need to use the following self-energy matrix, which contains four different combinations of the time arguments $(t_1,t_2)$ on the two branches of the closed contour:
\begin{align}
\Sigma^{(2,0)}_{p\sigma}(b,a)&=-\frac{2g_i^2}{V^2}e^{-\eta(|b|+|a|)} \sum_{\vec{kqr}}\delta_{\vec{p+k,q+r}}\notag \\
&\times\sum_{\alpha\beta\gamma}\mathbf{\bar{\Sigma}^{(2)}}(b,a)\delta_{\sigma,\beta}\delta_{\alpha\gamma},
\end{align}
\begin{align}
\Sigma^{(1,1)}_{p\sigma}(b,a)&=-\frac{4}{V^2}\theta(a)e^{-\eta |b|} \sum_{\vec{kqr}}\delta_{\vec{p+k,q+r}}\notag \\
&\times \sum_{\alpha\beta\gamma}g_i g^{(1)}_f(\sigma,\alpha;\sigma,\alpha)
\mathbf{\bar{\Sigma}^{(2)}}(b,a)\delta_{\sigma,\beta}\delta_{\alpha,\gamma},\end{align}
\begin{align}
\Sigma^{(0,2)}_{p\sigma}(b,a)&=-\frac{2}{V^2}\theta(b)\theta(a) \sum_{\vec{kqr}}\delta_{\vec{p+k,q+r}}\notag \\
&\times \sum_{\alpha\beta\gamma}g^{(2)}_f(\sigma,\alpha;\beta,\gamma)\mathbf{\bar{\Sigma}^{(2)}}(b,a),\end{align}
where $\mathbf{\bar{\Sigma}^{(2)}}(b,a)$ is the following matrix:
\begin{align}
\mathbf{\bar{\Sigma}}(b,a)=\begin{pmatrix}\bar{\Sigma}^{(2,T)}(b,a)&\bar{\Sigma}^{(2,>)}(\bar{b},a)\\\bar{\Sigma}^{(2,<)}(b,\bar{a}) &\bar{\Sigma}^{(2,\tilde{T})}(\bar{b},\bar{a})\end{pmatrix}
\end{align}
where, for the sake of brevity, we have suppressed the explicit dependence of $\Sigma^{(2)}(b,a) = \Sigma^{(2)}_{kqr,\alpha\beta\gamma}(b,a)$ on the momentum and spin indices. Furthermore, we have introduced the following notation:
 \begin{align}\label{eq:g2}
 &g^{(2)}_f(\sigma,\alpha;\beta,\gamma)=\sum_{J_1, J_2=0,2,..}\sum_{M_1 M_2}g_{J_1}g_{J_2} \langle J_1 M_1 |FF\sigma\alpha\rangle \times\notag\\
 &\langle FF \beta\gamma|J_1 M_1 \rangle \langle J_2 M_2 |FF\sigma\alpha\rangle\langle FF \beta\gamma|J_2 M_2 \rangle,
\end{align}
which  contains the information of coupling strength and Clebsch-Gordon coefficients. Using the diagram in Fig.~\ref{fig:feyn1} and free propagators in Eqs.~\eqref{eq:g1} to \eqref{eq:g4}, we can write down the following expression for the elements in the matrix:
\begin{align}
 \bar{\Sigma}^{(2,<)}(t_2,\bar{t}_1)&=i^3G^{<}_{k\alpha}(t_2,\bar{t}_1)G^{>}_{q\beta}(\bar{t}_1,t_2)G^{>}_{r\gamma}(\bar{t}_1,t_2),\notag\\
&=(1-n^0_{q\beta})(1-n^0_{r\gamma}) n^0_{k\alpha}e^{i(t_1-t_2)(\epsilon_{q}+\epsilon_{r}-\epsilon_{k})} ,\label{s1}\\
\bar{\Sigma}^{(2,>)}(\bar{t}_2,t_1)&=i^3G^{>}_{k\alpha}(\bar{t}_2,t_1)G^{<}_{q\beta}(t_1,\bar{t}_2)G^{<}_{r\gamma}(t_1,\bar{t}_2),\notag \\
  &=n^0_{q\beta}n^0_{r\gamma}(1-n^0_{k\alpha})e^{i(t_1-t_2)(\epsilon_{q}+\epsilon_{r}-\epsilon_{k})},\label{s2}\\        
 \bar{\Sigma}^{(2,T)}(t_2,t_1)&=i^3G^{T}_{k\alpha}(t_2,t_1)G^{T}_{q\beta}(t_1,t_2)G^T_{r\gamma}(t_1,t_2),\notag\\
 =\theta(t_2&-t_1)\Sigma^{(2,>)}(t_2,\bar{t}_1)+\theta(t_1-t_2)\Sigma^{(2,>)}(\bar{t}_2,t_1),\label{s3} \\ 
 \bar{\Sigma}^{(2,\tilde{T})}(\bar{t}_2,\bar{t}_1)&=i^3G^{\bar{T}}_{k\alpha}(\bar{t}_2,\bar{t}_1)G^{\bar{T}}_{q\beta}(\bar{t}_1,\bar{t}_2)G^{\bar{T}}_{r\gamma}(\bar{t}_1,\bar{t}_2),\notag\\
=\theta(t_1&-t_2)\Sigma^{(2,>)}(t_2,\bar{t}_1)+\theta(t_2-t_1)\Sigma^{(2,>)}(\bar{t}_2,t_1).\label{s4}
\end{align}
\section{Evolution of the spin population}\label{app:b}

Using the result of momentum distribution, we can find the change in populations to leading order:
\begin{align}
\delta N_{\sigma}&=\sum_{\vec{p}}n^{(2)}_{p\sigma}(t)\\ 
&=-\frac{2}{V^2}\sum_{pkqr}\sum_{\sigma\alpha\beta\gamma}Q^{\sigma\alpha\beta\gamma}_{pkqr} g^{(2)}(\sigma,\alpha;\beta,\gamma) F(E_{pkqr},t).
\end{align}
Notice that this result does not involve the corrections of $O(U^2_0)$ and $O(U_1 U_0)$ to
the momentum distribution. This is because the 
initial interaction conserves the population of the different spin components.

For intermediate to long times,  we notice that, formally,
\begin{equation}
 \lim_{t\to +\infty} F(E,t)\propto t\, \delta(E)
 \end{equation}
and therefore the rate of  population change of the different components is  given by  golden-rule expression in Eq.~\eqref{eq:N}. This rate is proportional to the phase-space volume available for scattering with SCCs. We note, this is result is independent of the initial interaction which preserves the SU$(N)$ symmetry.
\bibliography{cite}

\begin{thebibliography}{86}%
\makeatletter
\providecommand \@ifxundefined [1]{%
 \@ifx{#1\undefined}
}%
\providecommand \@ifnum [1]{%
 \ifnum #1\expandafter \@firstoftwo
 \else \expandafter \@secondoftwo
 \fi
}%
\providecommand \@ifx [1]{%
 \ifx #1\expandafter \@firstoftwo
 \else \expandafter \@secondoftwo
 \fi
}%
\providecommand \natexlab [1]{#1}%
\providecommand \enquote  [1]{``#1''}%
\providecommand \bibnamefont  [1]{#1}%
\providecommand \bibfnamefont [1]{#1}%
\providecommand \citenamefont [1]{#1}%
\providecommand \href@noop [0]{\@secondoftwo}%
\providecommand \href [0]{\begingroup \@sanitize@url \@href}%
\providecommand \@href[1]{\@@startlink{#1}\@@href}%
\providecommand \@@href[1]{\endgroup#1\@@endlink}%
\providecommand \@sanitize@url [0]{\catcode `\\12\catcode `\$12\catcode
  `\&12\catcode `\#12\catcode `\^12\catcode `\_12\catcode `\%12\relax}%
\providecommand \@@startlink[1]{}%
\providecommand \@@endlink[0]{}%
\providecommand \url  [0]{\begingroup\@sanitize@url \@url }%
\providecommand \@url [1]{\endgroup\@href {#1}{\urlprefix }}%
\providecommand \urlprefix  [0]{URL }%
\providecommand \Eprint [0]{\href }%
\providecommand \doibase [0]{http://dx.doi.org/}%
\providecommand \selectlanguage [0]{\@gobble}%
\providecommand \bibinfo  [0]{\@secondoftwo}%
\providecommand \bibfield  [0]{\@secondoftwo}%
\providecommand \translation [1]{[#1]}%
\providecommand \BibitemOpen [0]{}%
\providecommand \bibitemStop [0]{}%
\providecommand \bibitemNoStop [0]{.\EOS\space}%
\providecommand \EOS [0]{\spacefactor3000\relax}%
\providecommand \BibitemShut  [1]{\csname bibitem#1\endcsname}%
\let\auto@bib@innerbib\@empty
\bibitem [{\citenamefont {Cazalilla}\ \emph {et~al.}(2009)\citenamefont
  {Cazalilla}, \citenamefont {Ho},\ and\ \citenamefont {Ueda}}]{Miguel2009}%
  \BibitemOpen
  \bibfield  {author} {\bibinfo {author} {\bibfnamefont {M.~A.}\ \bibnamefont
  {Cazalilla}}, \bibinfo {author} {\bibfnamefont {A.~F.}\ \bibnamefont {Ho}}, \
  and\ \bibinfo {author} {\bibfnamefont {M.}~\bibnamefont {Ueda}},\ }\href
  {http://stacks.iop.org/1367-2630/11/i=10/a=103033} {\bibfield  {journal}
  {\bibinfo  {journal} {New Journal of Physics}\ }\textbf {\bibinfo {volume}
  {11}},\ \bibinfo {pages} {103033} (\bibinfo {year} {2009})}\BibitemShut
  {NoStop}%
\bibitem [{\citenamefont {Gorshkov}\ \emph {et~al.}(2010)\citenamefont
  {Gorshkov}, \citenamefont {Hermele}, \citenamefont {Gurarie}, \citenamefont
  {Xu}, \citenamefont {Julienne}, \citenamefont {Ye}, \citenamefont {Zoller},
  \citenamefont {Demler}, \citenamefont {Lukin},\ and\ \citenamefont
  {Rey}}]{Gorshkov2010}%
  \BibitemOpen
  \bibfield  {author} {\bibinfo {author} {\bibfnamefont {A.~V.}\ \bibnamefont
  {Gorshkov}}, \bibinfo {author} {\bibfnamefont {M.}~\bibnamefont {Hermele}},
  \bibinfo {author} {\bibfnamefont {V.}~\bibnamefont {Gurarie}}, \bibinfo
  {author} {\bibfnamefont {C.}~\bibnamefont {Xu}}, \bibinfo {author}
  {\bibfnamefont {P.~S.}\ \bibnamefont {Julienne}}, \bibinfo {author}
  {\bibfnamefont {J.}~\bibnamefont {Ye}}, \bibinfo {author} {\bibfnamefont
  {P.}~\bibnamefont {Zoller}}, \bibinfo {author} {\bibfnamefont
  {E.}~\bibnamefont {Demler}}, \bibinfo {author} {\bibfnamefont {M.~D.}\
  \bibnamefont {Lukin}}, \ and\ \bibinfo {author} {\bibfnamefont {A.~M.}\
  \bibnamefont {Rey}},\ }\href {http://dx.doi.org/10.1038/nphys1535} {\bibfield
   {journal} {\bibinfo  {journal} {Nat Phys}\ }\textbf {\bibinfo {volume}
  {6}},\ \bibinfo {pages} {289} (\bibinfo {year} {2010})}\BibitemShut {NoStop}%
\bibitem [{\citenamefont {Cazalilla}\ and\ \citenamefont
  {Rey}(2014)}]{Miguel2014}%
  \BibitemOpen
  \bibfield  {author} {\bibinfo {author} {\bibfnamefont {M.~A.}\ \bibnamefont
  {Cazalilla}}\ and\ \bibinfo {author} {\bibfnamefont {A.~M.}\ \bibnamefont
  {Rey}},\ }\href {http://stacks.iop.org/0034-4885/77/i=12/a=124401} {\bibfield
   {journal} {\bibinfo  {journal} {Reports on Progress in Physics}\ }\textbf
  {\bibinfo {volume} {77}},\ \bibinfo {pages} {124401} (\bibinfo {year}
  {2014})}\BibitemShut {NoStop}%
\bibitem [{\citenamefont {Yamazaki}\ \emph
  {et~al.}(2010{\natexlab{a}})\citenamefont {Yamazaki}, \citenamefont {Taie},
  \citenamefont {Sugawa},\ and\ \citenamefont {Takahashi}}]{Yamazaki2010}%
  \BibitemOpen
  \bibfield  {author} {\bibinfo {author} {\bibfnamefont {R.}~\bibnamefont
  {Yamazaki}}, \bibinfo {author} {\bibfnamefont {S.}~\bibnamefont {Taie}},
  \bibinfo {author} {\bibfnamefont {S.}~\bibnamefont {Sugawa}}, \ and\ \bibinfo
  {author} {\bibfnamefont {Y.}~\bibnamefont {Takahashi}},\ }\href {\doibase
  10.1103/PhysRevLett.105.050405} {\bibfield  {journal} {\bibinfo  {journal}
  {Phys. Rev. Lett.}\ }\textbf {\bibinfo {volume} {105}},\ \bibinfo {pages}
  {050405} (\bibinfo {year} {2010}{\natexlab{a}})}\BibitemShut {NoStop}%
\bibitem [{\citenamefont {Stellmer}\ \emph {et~al.}(2011)\citenamefont
  {Stellmer}, \citenamefont {Grimm},\ and\ \citenamefont
  {Schreck}}]{PhysRevA.84.043611_schreck}%
  \BibitemOpen
  \bibfield  {author} {\bibinfo {author} {\bibfnamefont {S.}~\bibnamefont
  {Stellmer}}, \bibinfo {author} {\bibfnamefont {R.}~\bibnamefont {Grimm}}, \
  and\ \bibinfo {author} {\bibfnamefont {F.}~\bibnamefont {Schreck}},\ }\href
  {\doibase 10.1103/PhysRevA.84.043611} {\bibfield  {journal} {\bibinfo
  {journal} {Phys. Rev. A}\ }\textbf {\bibinfo {volume} {84}},\ \bibinfo
  {pages} {043611} (\bibinfo {year} {2011})}\BibitemShut {NoStop}%
\bibitem [{\citenamefont {Affleck}\ and\ \citenamefont
  {Marston}(1988)}]{Affleck1988}%
  \BibitemOpen
  \bibfield  {author} {\bibinfo {author} {\bibfnamefont {I.}~\bibnamefont
  {Affleck}}\ and\ \bibinfo {author} {\bibfnamefont {J.~B.}\ \bibnamefont
  {Marston}},\ }\href {\doibase 10.1103/PhysRevB.37.3774} {\bibfield  {journal}
  {\bibinfo  {journal} {Phys. Rev. B}\ }\textbf {\bibinfo {volume} {37}},\
  \bibinfo {pages} {3774} (\bibinfo {year} {1988})}\BibitemShut {NoStop}%
\bibitem [{\citenamefont {Marston}\ and\ \citenamefont
  {Affleck}(1989)}]{Marston1989}%
  \BibitemOpen
  \bibfield  {author} {\bibinfo {author} {\bibfnamefont {J.~B.}\ \bibnamefont
  {Marston}}\ and\ \bibinfo {author} {\bibfnamefont {I.}~\bibnamefont
  {Affleck}},\ }\href {\doibase 10.1103/PhysRevB.39.11538} {\bibfield
  {journal} {\bibinfo  {journal} {Phys. Rev. B}\ }\textbf {\bibinfo {volume}
  {39}},\ \bibinfo {pages} {11538} (\bibinfo {year} {1989})}\BibitemShut
  {NoStop}%
\bibitem [{\citenamefont {Coleman}(1983)}]{PhysRevB.28.5255}%
  \BibitemOpen
  \bibfield  {author} {\bibinfo {author} {\bibfnamefont {P.}~\bibnamefont
  {Coleman}},\ }\href {\doibase 10.1103/PhysRevB.28.5255} {\bibfield  {journal}
  {\bibinfo  {journal} {Phys. Rev. B}\ }\textbf {\bibinfo {volume} {28}},\
  \bibinfo {pages} {5255} (\bibinfo {year} {1983})}\BibitemShut {NoStop}%
\bibitem [{\citenamefont {Read}\ and\ \citenamefont {Newns}(1983)}]{Read1983}%
  \BibitemOpen
  \bibfield  {author} {\bibinfo {author} {\bibfnamefont {N.}~\bibnamefont
  {Read}}\ and\ \bibinfo {author} {\bibfnamefont {D.~M.}\ \bibnamefont
  {Newns}},\ }\href {http://stacks.iop.org/0022-3719/16/i=29/a=007} {\bibfield
  {journal} {\bibinfo  {journal} {Journal of Physics C: Solid State Physics}\
  }\textbf {\bibinfo {volume} {16}},\ \bibinfo {pages} {L1055} (\bibinfo {year}
  {1983})}\BibitemShut {NoStop}%
\bibitem [{\citenamefont {Read}\ and\ \citenamefont {Newns}(1987)}]{Read1987}%
  \BibitemOpen
  \bibfield  {author} {\bibinfo {author} {\bibfnamefont {N.}~\bibnamefont
  {Read}}\ and\ \bibinfo {author} {\bibfnamefont {D.}~\bibnamefont {Newns}},\
  }\href@noop {} {\bibfield  {journal} {\bibinfo  {journal} {Adv. Phys.}\
  }\textbf {\bibinfo {volume} {36}} (\bibinfo {year} {1987})}\BibitemShut
  {NoStop}%
\bibitem [{\citenamefont {Read}\ and\ \citenamefont
  {Sachdev}(1989)}]{Read1989}%
  \BibitemOpen
  \bibfield  {author} {\bibinfo {author} {\bibfnamefont {N.}~\bibnamefont
  {Read}}\ and\ \bibinfo {author} {\bibfnamefont {S.}~\bibnamefont {Sachdev}},\
  }\href@noop {} {\bibfield  {journal} {\bibinfo  {journal} {Nucl. Phys. B}\
  }\textbf {\bibinfo {volume} {316}} (\bibinfo {year} {1989})}\BibitemShut
  {NoStop}%
\bibitem [{\citenamefont {Cheng}\ and\ \citenamefont
  {Yip}(2017)}]{PhysRevA.95.033619}%
  \BibitemOpen
  \bibfield  {author} {\bibinfo {author} {\bibfnamefont {C.-H.}\ \bibnamefont
  {Cheng}}\ and\ \bibinfo {author} {\bibfnamefont {S.-K.}\ \bibnamefont
  {Yip}},\ }\href {\doibase 10.1103/PhysRevA.95.033619} {\bibfield  {journal}
  {\bibinfo  {journal} {Phys. Rev. A}\ }\textbf {\bibinfo {volume} {95}},\
  \bibinfo {pages} {033619} (\bibinfo {year} {2017})}\BibitemShut {NoStop}%
\bibitem [{\citenamefont {Barbarino}\ \emph {et~al.}(2015)\citenamefont
  {Barbarino}, \citenamefont {Taddia}, \citenamefont {Rossini}, \citenamefont
  {Mazza},\ and\ \citenamefont {Fazio}}]{Barbarino2015}%
  \BibitemOpen
  \bibfield  {author} {\bibinfo {author} {\bibfnamefont {S.}~\bibnamefont
  {Barbarino}}, \bibinfo {author} {\bibfnamefont {L.}~\bibnamefont {Taddia}},
  \bibinfo {author} {\bibfnamefont {D.}~\bibnamefont {Rossini}}, \bibinfo
  {author} {\bibfnamefont {L.}~\bibnamefont {Mazza}}, \ and\ \bibinfo {author}
  {\bibfnamefont {R.}~\bibnamefont {Fazio}},\ }\href@noop {} {\bibfield
  {journal} {\bibinfo  {journal} {Nat. Commun.}\ }\textbf {\bibinfo {volume}
  {6}} (\bibinfo {year} {2015})}\BibitemShut {NoStop}%
\bibitem [{\citenamefont {Barbarino}\ \emph {et~al.}(2016)\citenamefont
  {Barbarino}, \citenamefont {Taddia}, \citenamefont {Rossini}, \citenamefont
  {Mazza},\ and\ \citenamefont {Fazio}}]{Barbarino2016}%
  \BibitemOpen
  \bibfield  {author} {\bibinfo {author} {\bibfnamefont {S.}~\bibnamefont
  {Barbarino}}, \bibinfo {author} {\bibfnamefont {L.}~\bibnamefont {Taddia}},
  \bibinfo {author} {\bibfnamefont {D.}~\bibnamefont {Rossini}}, \bibinfo
  {author} {\bibfnamefont {L.}~\bibnamefont {Mazza}}, \ and\ \bibinfo {author}
  {\bibfnamefont {R.}~\bibnamefont {Fazio}},\ }\href
  {http://stacks.iop.org/1367-2630/18/i=3/a=035010} {\bibfield  {journal}
  {\bibinfo  {journal} {New Journal of Physics}\ }\textbf {\bibinfo {volume}
  {18}},\ \bibinfo {pages} {035010} (\bibinfo {year} {2016})}\BibitemShut
  {NoStop}%
\bibitem [{\citenamefont {Celi}\ \emph {et~al.}(2014)\citenamefont {Celi},
  \citenamefont {Massignan}, \citenamefont {Ruseckas}, \citenamefont {Goldman},
  \citenamefont {Spielman}, \citenamefont {Juzeli\ifmmode~\bar{u}\else
  \={u}\fi{}nas},\ and\ \citenamefont {Lewenstein}}]{PhysRevLett.112.043001}%
  \BibitemOpen
  \bibfield  {author} {\bibinfo {author} {\bibfnamefont {A.}~\bibnamefont
  {Celi}}, \bibinfo {author} {\bibfnamefont {P.}~\bibnamefont {Massignan}},
  \bibinfo {author} {\bibfnamefont {J.}~\bibnamefont {Ruseckas}}, \bibinfo
  {author} {\bibfnamefont {N.}~\bibnamefont {Goldman}}, \bibinfo {author}
  {\bibfnamefont {I.~B.}\ \bibnamefont {Spielman}}, \bibinfo {author}
  {\bibfnamefont {G.}~\bibnamefont {Juzeli\ifmmode~\bar{u}\else
  \={u}\fi{}nas}}, \ and\ \bibinfo {author} {\bibfnamefont {M.}~\bibnamefont
  {Lewenstein}},\ }\href {\doibase 10.1103/PhysRevLett.112.043001} {\bibfield
  {journal} {\bibinfo  {journal} {Phys. Rev. Lett.}\ }\textbf {\bibinfo
  {volume} {112}},\ \bibinfo {pages} {043001} (\bibinfo {year}
  {2014})}\BibitemShut {NoStop}%
\bibitem [{\citenamefont {Beverland}\ \emph {et~al.}(2016)\citenamefont
  {Beverland}, \citenamefont {Alagic}, \citenamefont {Martin}, \citenamefont
  {Koller}, \citenamefont {Rey},\ and\ \citenamefont
  {Gorshkov}}]{PhysRevA.93.051601}%
  \BibitemOpen
  \bibfield  {author} {\bibinfo {author} {\bibfnamefont {M.~E.}\ \bibnamefont
  {Beverland}}, \bibinfo {author} {\bibfnamefont {G.}~\bibnamefont {Alagic}},
  \bibinfo {author} {\bibfnamefont {M.~J.}\ \bibnamefont {Martin}}, \bibinfo
  {author} {\bibfnamefont {A.~P.}\ \bibnamefont {Koller}}, \bibinfo {author}
  {\bibfnamefont {A.~M.}\ \bibnamefont {Rey}}, \ and\ \bibinfo {author}
  {\bibfnamefont {A.~V.}\ \bibnamefont {Gorshkov}},\ }\href {\doibase
  10.1103/PhysRevA.93.051601} {\bibfield  {journal} {\bibinfo  {journal} {Phys.
  Rev. A}\ }\textbf {\bibinfo {volume} {93}},\ \bibinfo {pages} {051601}
  (\bibinfo {year} {2016})}\BibitemShut {NoStop}%
\bibitem [{\citenamefont {Hermele}\ \emph {et~al.}(2009)\citenamefont
  {Hermele}, \citenamefont {Gurarie},\ and\ \citenamefont
  {Rey}}]{Hermele_magnetism}%
  \BibitemOpen
  \bibfield  {author} {\bibinfo {author} {\bibfnamefont {M.}~\bibnamefont
  {Hermele}}, \bibinfo {author} {\bibfnamefont {V.}~\bibnamefont {Gurarie}}, \
  and\ \bibinfo {author} {\bibfnamefont {A.~M.}\ \bibnamefont {Rey}},\ }\href
  {\doibase 10.1103/PhysRevLett.103.135301} {\bibfield  {journal} {\bibinfo
  {journal} {Phys. Rev. Lett.}\ }\textbf {\bibinfo {volume} {103}},\ \bibinfo
  {pages} {135301} (\bibinfo {year} {2009})}\BibitemShut {NoStop}%
\bibitem [{\citenamefont {Dufour}\ and\ \citenamefont
  {Mila}(2016)}]{Mila_chiral}%
  \BibitemOpen
  \bibfield  {author} {\bibinfo {author} {\bibfnamefont {J.}~\bibnamefont
  {Dufour}}\ and\ \bibinfo {author} {\bibfnamefont {F.}~\bibnamefont {Mila}},\
  }\href {\doibase 10.1103/PhysRevA.94.033617} {\bibfield  {journal} {\bibinfo
  {journal} {Phys. Rev. A}\ }\textbf {\bibinfo {volume} {94}},\ \bibinfo
  {pages} {033617} (\bibinfo {year} {2016})}\BibitemShut {NoStop}%
\bibitem [{\citenamefont {Nataf}\ and\ \citenamefont {Mila}(2016)}]{Mila_sun}%
  \BibitemOpen
  \bibfield  {author} {\bibinfo {author} {\bibfnamefont {P.}~\bibnamefont
  {Nataf}}\ and\ \bibinfo {author} {\bibfnamefont {F.}~\bibnamefont {Mila}},\
  }\href {\doibase 10.1103/PhysRevB.93.155134} {\bibfield  {journal} {\bibinfo
  {journal} {Phys. Rev. B}\ }\textbf {\bibinfo {volume} {93}},\ \bibinfo
  {pages} {155134} (\bibinfo {year} {2016})}\BibitemShut {NoStop}%
\bibitem [{\citenamefont {Sugawa}\ \emph {et~al.}(2011)\citenamefont {Sugawa},
  \citenamefont {Inaba}, \citenamefont {Taie}, \citenamefont {Yamazaki},
  \citenamefont {Yamashita},\ and\ \citenamefont
  {Takahashi}}]{Sugawa:2011aa_SUnexp}%
  \BibitemOpen
  \bibfield  {author} {\bibinfo {author} {\bibfnamefont {S.}~\bibnamefont
  {Sugawa}}, \bibinfo {author} {\bibfnamefont {K.}~\bibnamefont {Inaba}},
  \bibinfo {author} {\bibfnamefont {S.}~\bibnamefont {Taie}}, \bibinfo {author}
  {\bibfnamefont {R.}~\bibnamefont {Yamazaki}}, \bibinfo {author}
  {\bibfnamefont {M.}~\bibnamefont {Yamashita}}, \ and\ \bibinfo {author}
  {\bibfnamefont {Y.}~\bibnamefont {Takahashi}},\ }\href
  {https://doi.org/10.1038/nphys2028} {\bibfield  {journal} {\bibinfo
  {journal} {Nature Physics}\ }\textbf {\bibinfo {volume} {7}},\ \bibinfo
  {pages} {642 EP } (\bibinfo {year} {2011})}\BibitemShut {NoStop}%
\bibitem [{\citenamefont {Taie}\ \emph {et~al.}(2012)\citenamefont {Taie},
  \citenamefont {Yamazaki}, \citenamefont {Sugawa},\ and\ \citenamefont
  {Takahashi}}]{Taie:2012aa_SUnexp}%
  \BibitemOpen
  \bibfield  {author} {\bibinfo {author} {\bibfnamefont {S.}~\bibnamefont
  {Taie}}, \bibinfo {author} {\bibfnamefont {R.}~\bibnamefont {Yamazaki}},
  \bibinfo {author} {\bibfnamefont {S.}~\bibnamefont {Sugawa}}, \ and\ \bibinfo
  {author} {\bibfnamefont {Y.}~\bibnamefont {Takahashi}},\ }\href
  {https://doi.org/10.1038/nphys2430} {\bibfield  {journal} {\bibinfo
  {journal} {Nature Physics}\ }\textbf {\bibinfo {volume} {8}},\ \bibinfo
  {pages} {825 EP } (\bibinfo {year} {2012})}\BibitemShut {NoStop}%
\bibitem [{\citenamefont {Ozawa}\ \emph {et~al.}(2018)\citenamefont {Ozawa},
  \citenamefont {Taie}, \citenamefont {Takasu},\ and\ \citenamefont
  {Takahashi}}]{PhysRevLett.121.225303_SUnexp}%
  \BibitemOpen
  \bibfield  {author} {\bibinfo {author} {\bibfnamefont {H.}~\bibnamefont
  {Ozawa}}, \bibinfo {author} {\bibfnamefont {S.}~\bibnamefont {Taie}},
  \bibinfo {author} {\bibfnamefont {Y.}~\bibnamefont {Takasu}}, \ and\ \bibinfo
  {author} {\bibfnamefont {Y.}~\bibnamefont {Takahashi}},\ }\href {\doibase
  10.1103/PhysRevLett.121.225303} {\bibfield  {journal} {\bibinfo  {journal}
  {Phys. Rev. Lett.}\ }\textbf {\bibinfo {volume} {121}},\ \bibinfo {pages}
  {225303} (\bibinfo {year} {2018})}\BibitemShut {NoStop}%
\bibitem [{\citenamefont {Hofrichter}\ \emph {et~al.}(2016)\citenamefont
  {Hofrichter}, \citenamefont {Riegger}, \citenamefont {Scazza}, \citenamefont
  {H\"ofer}, \citenamefont {Fernandes}, \citenamefont {Bloch},\ and\
  \citenamefont {F\"olling}}]{PhysRevX.6.021030_SUnexp}%
  \BibitemOpen
  \bibfield  {author} {\bibinfo {author} {\bibfnamefont {C.}~\bibnamefont
  {Hofrichter}}, \bibinfo {author} {\bibfnamefont {L.}~\bibnamefont {Riegger}},
  \bibinfo {author} {\bibfnamefont {F.}~\bibnamefont {Scazza}}, \bibinfo
  {author} {\bibfnamefont {M.}~\bibnamefont {H\"ofer}}, \bibinfo {author}
  {\bibfnamefont {D.~R.}\ \bibnamefont {Fernandes}}, \bibinfo {author}
  {\bibfnamefont {I.}~\bibnamefont {Bloch}}, \ and\ \bibinfo {author}
  {\bibfnamefont {S.}~\bibnamefont {F\"olling}},\ }\href {\doibase
  10.1103/PhysRevX.6.021030} {\bibfield  {journal} {\bibinfo  {journal} {Phys.
  Rev. X}\ }\textbf {\bibinfo {volume} {6}},\ \bibinfo {pages} {021030}
  (\bibinfo {year} {2016})}\BibitemShut {NoStop}%
\bibitem [{\citenamefont {Zhang}\ \emph {et~al.}(2014)\citenamefont {Zhang},
  \citenamefont {Bishof}, \citenamefont {Bromley}, \citenamefont {Kraus},
  \citenamefont {Safronova}, \citenamefont {Zoller}, \citenamefont {Rey},\ and\
  \citenamefont {Ye}}]{Zhang2014}%
  \BibitemOpen
  \bibfield  {author} {\bibinfo {author} {\bibfnamefont {X.}~\bibnamefont
  {Zhang}}, \bibinfo {author} {\bibfnamefont {M.}~\bibnamefont {Bishof}},
  \bibinfo {author} {\bibfnamefont {S.~L.}\ \bibnamefont {Bromley}}, \bibinfo
  {author} {\bibfnamefont {C.~V.}\ \bibnamefont {Kraus}}, \bibinfo {author}
  {\bibfnamefont {M.~S.}\ \bibnamefont {Safronova}}, \bibinfo {author}
  {\bibfnamefont {P.}~\bibnamefont {Zoller}}, \bibinfo {author} {\bibfnamefont
  {A.~M.}\ \bibnamefont {Rey}}, \ and\ \bibinfo {author} {\bibfnamefont
  {J.}~\bibnamefont {Ye}},\ }\href {\doibase 10.1126/science.1254978}
  {\bibfield  {journal} {\bibinfo  {journal} {Science}\ }\textbf {\bibinfo
  {volume} {345}},\ \bibinfo {pages} {1467} (\bibinfo {year} {2014})},\ \Eprint
  {http://arxiv.org/abs/http://science.sciencemag.org/content/345/6203/1467.full.pdf}
  {http://science.sciencemag.org/content/345/6203/1467.full.pdf} \BibitemShut
  {NoStop}%
\bibitem [{\citenamefont {Cappellini}\ \emph {et~al.}(2014)\citenamefont
  {Cappellini}, \citenamefont {Mancini}, \citenamefont {Pagano}, \citenamefont
  {Lombardi}, \citenamefont {Livi}, \citenamefont {Siciliani~de Cumis},
  \citenamefont {Cancio}, \citenamefont {Pizzocaro}, \citenamefont {Calonico},
  \citenamefont {Levi}, \citenamefont {Sias}, \citenamefont {Catani},
  \citenamefont {Inguscio},\ and\ \citenamefont {Fallani}}]{Cappellini2014}%
  \BibitemOpen
  \bibfield  {author} {\bibinfo {author} {\bibfnamefont {G.}~\bibnamefont
  {Cappellini}}, \bibinfo {author} {\bibfnamefont {M.}~\bibnamefont {Mancini}},
  \bibinfo {author} {\bibfnamefont {G.}~\bibnamefont {Pagano}}, \bibinfo
  {author} {\bibfnamefont {P.}~\bibnamefont {Lombardi}}, \bibinfo {author}
  {\bibfnamefont {L.}~\bibnamefont {Livi}}, \bibinfo {author} {\bibfnamefont
  {M.}~\bibnamefont {Siciliani~de Cumis}}, \bibinfo {author} {\bibfnamefont
  {P.}~\bibnamefont {Cancio}}, \bibinfo {author} {\bibfnamefont
  {M.}~\bibnamefont {Pizzocaro}}, \bibinfo {author} {\bibfnamefont
  {D.}~\bibnamefont {Calonico}}, \bibinfo {author} {\bibfnamefont
  {F.}~\bibnamefont {Levi}}, \bibinfo {author} {\bibfnamefont {C.}~\bibnamefont
  {Sias}}, \bibinfo {author} {\bibfnamefont {J.}~\bibnamefont {Catani}},
  \bibinfo {author} {\bibfnamefont {M.}~\bibnamefont {Inguscio}}, \ and\
  \bibinfo {author} {\bibfnamefont {L.}~\bibnamefont {Fallani}},\ }\href
  {\doibase 10.1103/PhysRevLett.113.120402} {\bibfield  {journal} {\bibinfo
  {journal} {Phys. Rev. Lett.}\ }\textbf {\bibinfo {volume} {113}},\ \bibinfo
  {pages} {120402} (\bibinfo {year} {2014})}\BibitemShut {NoStop}%
\bibitem [{\citenamefont {Scazza}\ \emph {et~al.}(2014)\citenamefont {Scazza},
  \citenamefont {Hofrichter}, \citenamefont {Hofer}, \citenamefont {De~Groot},
  \citenamefont {Bloch},\ and\ \citenamefont {Folling}}]{Scazza2014}%
  \BibitemOpen
  \bibfield  {author} {\bibinfo {author} {\bibfnamefont {F.}~\bibnamefont
  {Scazza}}, \bibinfo {author} {\bibfnamefont {C.}~\bibnamefont {Hofrichter}},
  \bibinfo {author} {\bibfnamefont {M.}~\bibnamefont {Hofer}}, \bibinfo
  {author} {\bibfnamefont {P.~C.}\ \bibnamefont {De~Groot}}, \bibinfo {author}
  {\bibfnamefont {I.}~\bibnamefont {Bloch}}, \ and\ \bibinfo {author}
  {\bibfnamefont {S.}~\bibnamefont {Folling}},\ }\href
  {http://dx.doi.org/10.1038/nphys3061} {\bibfield  {journal} {\bibinfo
  {journal} {Nat Phys}\ }\textbf {\bibinfo {volume} {10}},\ \bibinfo {pages}
  {779} (\bibinfo {year} {2014})}\BibitemShut {NoStop}%
\bibitem [{\citenamefont {Ozawa}\ and\ \citenamefont
  {Baym}(2010)}]{OzawaBaym2010}%
  \BibitemOpen
  \bibfield  {author} {\bibinfo {author} {\bibfnamefont {T.}~\bibnamefont
  {Ozawa}}\ and\ \bibinfo {author} {\bibfnamefont {G.}~\bibnamefont {Baym}},\
  }\href {\doibase 10.1103/PhysRevA.82.063615} {\bibfield  {journal} {\bibinfo
  {journal} {Phys. Rev. A}\ }\textbf {\bibinfo {volume} {82}},\ \bibinfo
  {pages} {063615} (\bibinfo {year} {2010})}\BibitemShut {NoStop}%
\bibitem [{\citenamefont {Banerjee}\ \emph {et~al.}(2013)\citenamefont
  {Banerjee}, \citenamefont {B\"ogli}, \citenamefont {Dalmonte}, \citenamefont
  {Rico}, \citenamefont {Stebler}, \citenamefont {Wiese},\ and\ \citenamefont
  {Zoller}}]{Banerjee2013}%
  \BibitemOpen
  \bibfield  {author} {\bibinfo {author} {\bibfnamefont {D.}~\bibnamefont
  {Banerjee}}, \bibinfo {author} {\bibfnamefont {M.}~\bibnamefont {B\"ogli}},
  \bibinfo {author} {\bibfnamefont {M.}~\bibnamefont {Dalmonte}}, \bibinfo
  {author} {\bibfnamefont {E.}~\bibnamefont {Rico}}, \bibinfo {author}
  {\bibfnamefont {P.}~\bibnamefont {Stebler}}, \bibinfo {author} {\bibfnamefont
  {U.-J.}\ \bibnamefont {Wiese}}, \ and\ \bibinfo {author} {\bibfnamefont
  {P.}~\bibnamefont {Zoller}},\ }\href {\doibase
  10.1103/PhysRevLett.110.125303} {\bibfield  {journal} {\bibinfo  {journal}
  {Phys. Rev. Lett.}\ }\textbf {\bibinfo {volume} {110}},\ \bibinfo {pages}
  {125303} (\bibinfo {year} {2013})}\BibitemShut {NoStop}%
\bibitem [{\citenamefont {Berges}\ \emph {et~al.}(2004)\citenamefont {Berges},
  \citenamefont {Bors\'anyi},\ and\ \citenamefont
  {Wetterich}}]{PhysRevLett.93.142002}%
  \BibitemOpen
  \bibfield  {author} {\bibinfo {author} {\bibfnamefont {J.}~\bibnamefont
  {Berges}}, \bibinfo {author} {\bibfnamefont {S.}~\bibnamefont {Bors\'anyi}},
  \ and\ \bibinfo {author} {\bibfnamefont {C.}~\bibnamefont {Wetterich}},\
  }\href {\doibase 10.1103/PhysRevLett.93.142002} {\bibfield  {journal}
  {\bibinfo  {journal} {Phys. Rev. Lett.}\ }\textbf {\bibinfo {volume} {93}},\
  \bibinfo {pages} {142002} (\bibinfo {year} {2004})}\BibitemShut {NoStop}%
\bibitem [{\citenamefont {Moeckel}\ and\ \citenamefont
  {Kehrein}(2008)}]{Moeckel2008}%
  \BibitemOpen
  \bibfield  {author} {\bibinfo {author} {\bibfnamefont {M.}~\bibnamefont
  {Moeckel}}\ and\ \bibinfo {author} {\bibfnamefont {S.}~\bibnamefont
  {Kehrein}},\ }\href {\doibase 10.1103/PhysRevLett.100.175702} {\bibfield
  {journal} {\bibinfo  {journal} {Phys. Rev. Lett.}\ }\textbf {\bibinfo
  {volume} {100}},\ \bibinfo {pages} {175702} (\bibinfo {year}
  {2008})}\BibitemShut {NoStop}%
\bibitem [{\citenamefont {Moeckel}\ and\ \citenamefont
  {Kehrein}(2009)}]{Moeckel2009}%
  \BibitemOpen
  \bibfield  {author} {\bibinfo {author} {\bibfnamefont {M.}~\bibnamefont
  {Moeckel}}\ and\ \bibinfo {author} {\bibfnamefont {S.}~\bibnamefont
  {Kehrein}},\ }\href {\doibase http://dx.doi.org/10.1016/j.aop.2009.03.009}
  {\bibfield  {journal} {\bibinfo  {journal} {Annals of Physics}\ }\textbf
  {\bibinfo {volume} {324}},\ \bibinfo {pages} {2146 } (\bibinfo {year}
  {2009})}\BibitemShut {NoStop}%
\bibitem [{\citenamefont {Nessi}\ \emph {et~al.}(2014)\citenamefont {Nessi},
  \citenamefont {Iucci},\ and\ \citenamefont
  {Cazalilla}}]{nessi_shorttime2014}%
  \BibitemOpen
  \bibfield  {author} {\bibinfo {author} {\bibfnamefont {N.}~\bibnamefont
  {Nessi}}, \bibinfo {author} {\bibfnamefont {A.}~\bibnamefont {Iucci}}, \ and\
  \bibinfo {author} {\bibfnamefont {M.~A.}\ \bibnamefont {Cazalilla}},\ }\href
  {\doibase 10.1103/PhysRevLett.113.210402} {\bibfield  {journal} {\bibinfo
  {journal} {Phys. Rev. Lett.}\ }\textbf {\bibinfo {volume} {113}},\ \bibinfo
  {pages} {210402} (\bibinfo {year} {2014})}\BibitemShut {NoStop}%
\bibitem [{\citenamefont {Hamerla}\ and\ \citenamefont
  {Uhrig}(2014)}]{nopreth2d_2014}%
  \BibitemOpen
  \bibfield  {author} {\bibinfo {author} {\bibfnamefont {S.~A.}\ \bibnamefont
  {Hamerla}}\ and\ \bibinfo {author} {\bibfnamefont {G.~S.}\ \bibnamefont
  {Uhrig}},\ }\href {\doibase 10.1103/PhysRevB.89.104301} {\bibfield  {journal}
  {\bibinfo  {journal} {Phys. Rev. B}\ }\textbf {\bibinfo {volume} {89}},\
  \bibinfo {pages} {104301} (\bibinfo {year} {2014})}\BibitemShut {NoStop}%
\bibitem [{\citenamefont {Eckstein}\ \emph {et~al.}(2009)\citenamefont
  {Eckstein}, \citenamefont {Kollar},\ and\ \citenamefont
  {Werner}}]{prethandth_2009_eckstein}%
  \BibitemOpen
  \bibfield  {author} {\bibinfo {author} {\bibfnamefont {M.}~\bibnamefont
  {Eckstein}}, \bibinfo {author} {\bibfnamefont {M.}~\bibnamefont {Kollar}}, \
  and\ \bibinfo {author} {\bibfnamefont {P.}~\bibnamefont {Werner}},\ }\href
  {\doibase 10.1103/PhysRevLett.103.056403} {\bibfield  {journal} {\bibinfo
  {journal} {Phys. Rev. Lett.}\ }\textbf {\bibinfo {volume} {103}},\ \bibinfo
  {pages} {056403} (\bibinfo {year} {2009})}\BibitemShut {NoStop}%
\bibitem [{\citenamefont {{Nessi}}\ and\ \citenamefont
  {{Iucci}}(2015)}]{Nessi_glass_2015}%
  \BibitemOpen
  \bibfield  {author} {\bibinfo {author} {\bibfnamefont {N.}~\bibnamefont
  {{Nessi}}}\ and\ \bibinfo {author} {\bibfnamefont {A.}~\bibnamefont
  {{Iucci}}},\ }\href@noop {} {\bibfield  {journal} {\bibinfo  {journal} {ArXiv
  e-prints}\ } (\bibinfo {year} {2015})},\ \Eprint
  {http://arxiv.org/abs/1503.02507} {arXiv:1503.02507 [cond-mat.quant-gas]}
  \BibitemShut {NoStop}%
\bibitem [{\citenamefont {Marcuzzi}\ \emph {et~al.}(2013)\citenamefont
  {Marcuzzi}, \citenamefont {Marino}, \citenamefont {Gambassi},\ and\
  \citenamefont {Silva}}]{prethandth_silva_2013}%
  \BibitemOpen
  \bibfield  {author} {\bibinfo {author} {\bibfnamefont {M.}~\bibnamefont
  {Marcuzzi}}, \bibinfo {author} {\bibfnamefont {J.}~\bibnamefont {Marino}},
  \bibinfo {author} {\bibfnamefont {A.}~\bibnamefont {Gambassi}}, \ and\
  \bibinfo {author} {\bibfnamefont {A.}~\bibnamefont {Silva}},\ }\href
  {\doibase 10.1103/PhysRevLett.111.197203} {\bibfield  {journal} {\bibinfo
  {journal} {Phys. Rev. Lett.}\ }\textbf {\bibinfo {volume} {111}},\ \bibinfo
  {pages} {197203} (\bibinfo {year} {2013})}\BibitemShut {NoStop}%
\bibitem [{\citenamefont {Essler}\ \emph {et~al.}(2014)\citenamefont {Essler},
  \citenamefont {Kehrein}, \citenamefont {Manmana},\ and\ \citenamefont
  {Robinson}}]{turnable_integrablity_2014}%
  \BibitemOpen
  \bibfield  {author} {\bibinfo {author} {\bibfnamefont {F.~H.~L.}\
  \bibnamefont {Essler}}, \bibinfo {author} {\bibfnamefont {S.}~\bibnamefont
  {Kehrein}}, \bibinfo {author} {\bibfnamefont {S.~R.}\ \bibnamefont
  {Manmana}}, \ and\ \bibinfo {author} {\bibfnamefont {N.~J.}\ \bibnamefont
  {Robinson}},\ }\href {\doibase 10.1103/PhysRevB.89.165104} {\bibfield
  {journal} {\bibinfo  {journal} {Phys. Rev. B}\ }\textbf {\bibinfo {volume}
  {89}},\ \bibinfo {pages} {165104} (\bibinfo {year} {2014})}\BibitemShut
  {NoStop}%
\bibitem [{\citenamefont {Langen}\ \emph {et~al.}(2016)\citenamefont {Langen},
  \citenamefont {Gasenzer},\ and\ \citenamefont
  {Schmiedmayer}}]{nearintegrable_Lagen2016}%
  \BibitemOpen
  \bibfield  {author} {\bibinfo {author} {\bibfnamefont {T.}~\bibnamefont
  {Langen}}, \bibinfo {author} {\bibfnamefont {T.}~\bibnamefont {Gasenzer}}, \
  and\ \bibinfo {author} {\bibfnamefont {J.}~\bibnamefont {Schmiedmayer}},\
  }\href {http://stacks.iop.org/1742-5468/2016/i=6/a=064009} {\bibfield
  {journal} {\bibinfo  {journal} {Journal of Statistical Mechanics: Theory and
  Experiment}\ }\textbf {\bibinfo {volume} {2016}},\ \bibinfo {pages} {064009}
  (\bibinfo {year} {2016})}\BibitemShut {NoStop}%
\bibitem [{\citenamefont {Moeckel}\ and\ \citenamefont
  {Kehrein}(2010)}]{Moeckel2010}%
  \BibitemOpen
  \bibfield  {author} {\bibinfo {author} {\bibfnamefont {M.}~\bibnamefont
  {Moeckel}}\ and\ \bibinfo {author} {\bibfnamefont {S.}~\bibnamefont
  {Kehrein}},\ }\href {http://stacks.iop.org/1367-2630/12/i=5/a=055016}
  {\bibfield  {journal} {\bibinfo  {journal} {New Journal of Physics}\ }\textbf
  {\bibinfo {volume} {12}},\ \bibinfo {pages} {055016} (\bibinfo {year}
  {2010})}\BibitemShut {NoStop}%
\bibitem [{\citenamefont {{Rademaker}}(2017)}]{spinprethermal2017}%
  \BibitemOpen
  \bibfield  {author} {\bibinfo {author} {\bibfnamefont {L.}~\bibnamefont
  {{Rademaker}}},\ }\href@noop {} {\bibfield  {journal} {\bibinfo  {journal}
  {ArXiv e-prints}\ } (\bibinfo {year} {2017})},\ \Eprint
  {http://arxiv.org/abs/1710.09761} {arXiv:1710.09761 [cond-mat.str-el]}
  \BibitemShut {NoStop}%
\bibitem [{\citenamefont {Mitra}(2013)}]{prethandth_mitra_2013}%
  \BibitemOpen
  \bibfield  {author} {\bibinfo {author} {\bibfnamefont {A.}~\bibnamefont
  {Mitra}},\ }\href {\doibase 10.1103/PhysRevB.87.205109} {\bibfield  {journal}
  {\bibinfo  {journal} {Phys. Rev. B}\ }\textbf {\bibinfo {volume} {87}},\
  \bibinfo {pages} {205109} (\bibinfo {year} {2013})}\BibitemShut {NoStop}%
\bibitem [{\citenamefont {Cazalilla}(2006)}]{PhysRevLett.97.156403}%
  \BibitemOpen
  \bibfield  {author} {\bibinfo {author} {\bibfnamefont {M.~A.}\ \bibnamefont
  {Cazalilla}},\ }\href {\doibase 10.1103/PhysRevLett.97.156403} {\bibfield
  {journal} {\bibinfo  {journal} {Phys. Rev. Lett.}\ }\textbf {\bibinfo
  {volume} {97}},\ \bibinfo {pages} {156403} (\bibinfo {year}
  {2006})}\BibitemShut {NoStop}%
\bibitem [{\citenamefont {Buchhold}\ \emph {et~al.}(2016)\citenamefont
  {Buchhold}, \citenamefont {Heyl},\ and\ \citenamefont
  {Diehl}}]{preth_th_luttinger_2016}%
  \BibitemOpen
  \bibfield  {author} {\bibinfo {author} {\bibfnamefont {M.}~\bibnamefont
  {Buchhold}}, \bibinfo {author} {\bibfnamefont {M.}~\bibnamefont {Heyl}}, \
  and\ \bibinfo {author} {\bibfnamefont {S.}~\bibnamefont {Diehl}},\ }\href
  {\doibase 10.1103/PhysRevA.94.013601} {\bibfield  {journal} {\bibinfo
  {journal} {Phys. Rev. A}\ }\textbf {\bibinfo {volume} {94}},\ \bibinfo
  {pages} {013601} (\bibinfo {year} {2016})}\BibitemShut {NoStop}%
\bibitem [{\citenamefont {Gong}\ and\ \citenamefont
  {Duan}(2013)}]{prethspinchain_Gong2013}%
  \BibitemOpen
  \bibfield  {author} {\bibinfo {author} {\bibfnamefont {Z.-X.}\ \bibnamefont
  {Gong}}\ and\ \bibinfo {author} {\bibfnamefont {L.-M.}\ \bibnamefont
  {Duan}},\ }\href {http://stacks.iop.org/1367-2630/15/i=11/a=113051}
  {\bibfield  {journal} {\bibinfo  {journal} {New Journal of Physics}\ }\textbf
  {\bibinfo {volume} {15}},\ \bibinfo {pages} {113051} (\bibinfo {year}
  {2013})}\BibitemShut {NoStop}%
\bibitem [{\citenamefont {Kollar}\ \emph {et~al.}(2011)\citenamefont {Kollar},
  \citenamefont {Wolf},\ and\ \citenamefont {Eckstein}}]{GGEpreth_2011_kollar}%
  \BibitemOpen
  \bibfield  {author} {\bibinfo {author} {\bibfnamefont {M.}~\bibnamefont
  {Kollar}}, \bibinfo {author} {\bibfnamefont {F.~A.}\ \bibnamefont {Wolf}}, \
  and\ \bibinfo {author} {\bibfnamefont {M.}~\bibnamefont {Eckstein}},\ }\href
  {\doibase 10.1103/PhysRevB.84.054304} {\bibfield  {journal} {\bibinfo
  {journal} {Phys. Rev. B}\ }\textbf {\bibinfo {volume} {84}},\ \bibinfo
  {pages} {054304} (\bibinfo {year} {2011})}\BibitemShut {NoStop}%
\bibitem [{\citenamefont {Cazalilla}\ and\ \citenamefont
  {Chung}(2016)}]{Miguel2016}%
  \BibitemOpen
  \bibfield  {author} {\bibinfo {author} {\bibfnamefont {M.~A.}\ \bibnamefont
  {Cazalilla}}\ and\ \bibinfo {author} {\bibfnamefont {M.-C.}\ \bibnamefont
  {Chung}},\ }\href {http://stacks.iop.org/1742-5468/2016/i=6/a=064004}
  {\bibfield  {journal} {\bibinfo  {journal} {Journal of Statistical Mechanics:
  Theory and Experiment}\ }\textbf {\bibinfo {volume} {2016}},\ \bibinfo
  {pages} {064004} (\bibinfo {year} {2016})}\BibitemShut {NoStop}%
\bibitem [{\citenamefont {Babadi}\ \emph {et~al.}(2015)\citenamefont {Babadi},
  \citenamefont {Demler},\ and\ \citenamefont {Knap}}]{spin_prethandth_2015}%
  \BibitemOpen
  \bibfield  {author} {\bibinfo {author} {\bibfnamefont {M.}~\bibnamefont
  {Babadi}}, \bibinfo {author} {\bibfnamefont {E.}~\bibnamefont {Demler}}, \
  and\ \bibinfo {author} {\bibfnamefont {M.}~\bibnamefont {Knap}},\ }\href
  {\doibase 10.1103/PhysRevX.5.041005} {\bibfield  {journal} {\bibinfo
  {journal} {Phys. Rev. X}\ }\textbf {\bibinfo {volume} {5}},\ \bibinfo {pages}
  {041005} (\bibinfo {year} {2015})}\BibitemShut {NoStop}%
\bibitem [{\citenamefont {{Alba}}\ and\ \citenamefont
  {{Fagotti}}(2017)}]{nearintegrable_alba_2017}%
  \BibitemOpen
  \bibfield  {author} {\bibinfo {author} {\bibfnamefont {V.}~\bibnamefont
  {{Alba}}}\ and\ \bibinfo {author} {\bibfnamefont {M.}~\bibnamefont
  {{Fagotti}}},\ }\href@noop {} {\bibfield  {journal} {\bibinfo  {journal}
  {ArXiv e-prints}\ } (\bibinfo {year} {2017})},\ \Eprint
  {http://arxiv.org/abs/1701.05552} {arXiv:1701.05552 [cond-mat.stat-mech]}
  \BibitemShut {NoStop}%
\bibitem [{\citenamefont {ping zou}\ and\ \citenamefont
  {Zhang}(2018)}]{preth_spin_short_2018}%
  \BibitemOpen
  \bibfield  {author} {\bibinfo {author} {\bibnamefont {ping zou}}\ and\
  \bibinfo {author} {\bibfnamefont {Z.-M.}\ \bibnamefont {Zhang}},\ }\href
  {http://iopscience.iop.org/10.1088/1361-6455/aac945} {\bibfield  {journal}
  {\bibinfo  {journal} {Journal of Physics B: Atomic, Molecular and Optical
  Physics}\ } (\bibinfo {year} {2018})}\BibitemShut {NoStop}%
\bibitem [{\citenamefont {Vidmar}\ and\ \citenamefont
  {Rigol}(2016)}]{rigol_gge}%
  \BibitemOpen
  \bibfield  {author} {\bibinfo {author} {\bibfnamefont {L.}~\bibnamefont
  {Vidmar}}\ and\ \bibinfo {author} {\bibfnamefont {M.}~\bibnamefont {Rigol}},\
  }\href {http://stacks.iop.org/1742-5468/2016/i=6/a=064007} {\bibfield
  {journal} {\bibinfo  {journal} {Journal of Statistical Mechanics: Theory and
  Experiment}\ }\textbf {\bibinfo {volume} {2016}},\ \bibinfo {pages} {064007}
  (\bibinfo {year} {2016})}\BibitemShut {NoStop}%
\bibitem [{\citenamefont {Rigol}\ \emph {et~al.}(2007)\citenamefont {Rigol},
  \citenamefont {Dunjko}, \citenamefont {Yurovsky},\ and\ \citenamefont
  {Olshanii}}]{integrable_rigol_2007}%
  \BibitemOpen
  \bibfield  {author} {\bibinfo {author} {\bibfnamefont {M.}~\bibnamefont
  {Rigol}}, \bibinfo {author} {\bibfnamefont {V.}~\bibnamefont {Dunjko}},
  \bibinfo {author} {\bibfnamefont {V.}~\bibnamefont {Yurovsky}}, \ and\
  \bibinfo {author} {\bibfnamefont {M.}~\bibnamefont {Olshanii}},\ }\href
  {\doibase 10.1103/PhysRevLett.98.050405} {\bibfield  {journal} {\bibinfo
  {journal} {Phys. Rev. Lett.}\ }\textbf {\bibinfo {volume} {98}},\ \bibinfo
  {pages} {050405} (\bibinfo {year} {2007})}\BibitemShut {NoStop}%
\bibitem [{\citenamefont {Calabrese}\ \emph {et~al.}(2011)\citenamefont
  {Calabrese}, \citenamefont {Essler},\ and\ \citenamefont
  {Fagotti}}]{isingchaing_2011}%
  \BibitemOpen
  \bibfield  {author} {\bibinfo {author} {\bibfnamefont {P.}~\bibnamefont
  {Calabrese}}, \bibinfo {author} {\bibfnamefont {F.~H.~L.}\ \bibnamefont
  {Essler}}, \ and\ \bibinfo {author} {\bibfnamefont {M.}~\bibnamefont
  {Fagotti}},\ }\href {\doibase 10.1103/PhysRevLett.106.227203} {\bibfield
  {journal} {\bibinfo  {journal} {Phys. Rev. Lett.}\ }\textbf {\bibinfo
  {volume} {106}},\ \bibinfo {pages} {227203} (\bibinfo {year}
  {2011})}\BibitemShut {NoStop}%
\bibitem [{\citenamefont {Ilievski}\ \emph {et~al.}(2015)\citenamefont
  {Ilievski}, \citenamefont {De~Nardis}, \citenamefont {Wouters}, \citenamefont
  {Caux}, \citenamefont {Essler},\ and\ \citenamefont {Prosen}}]{GGE_2015}%
  \BibitemOpen
  \bibfield  {author} {\bibinfo {author} {\bibfnamefont {E.}~\bibnamefont
  {Ilievski}}, \bibinfo {author} {\bibfnamefont {J.}~\bibnamefont {De~Nardis}},
  \bibinfo {author} {\bibfnamefont {B.}~\bibnamefont {Wouters}}, \bibinfo
  {author} {\bibfnamefont {J.-S.}\ \bibnamefont {Caux}}, \bibinfo {author}
  {\bibfnamefont {F.~H.~L.}\ \bibnamefont {Essler}}, \ and\ \bibinfo {author}
  {\bibfnamefont {T.}~\bibnamefont {Prosen}},\ }\href {\doibase
  10.1103/PhysRevLett.115.157201} {\bibfield  {journal} {\bibinfo  {journal}
  {Phys. Rev. Lett.}\ }\textbf {\bibinfo {volume} {115}},\ \bibinfo {pages}
  {157201} (\bibinfo {year} {2015})}\BibitemShut {NoStop}%
\bibitem [{\citenamefont {Wright}\ \emph {et~al.}(2014)\citenamefont {Wright},
  \citenamefont {Rigol}, \citenamefont {Davis},\ and\ \citenamefont
  {Kheruntsyan}}]{1dboson_2014}%
  \BibitemOpen
  \bibfield  {author} {\bibinfo {author} {\bibfnamefont {T.~M.}\ \bibnamefont
  {Wright}}, \bibinfo {author} {\bibfnamefont {M.}~\bibnamefont {Rigol}},
  \bibinfo {author} {\bibfnamefont {M.~J.}\ \bibnamefont {Davis}}, \ and\
  \bibinfo {author} {\bibfnamefont {K.~V.}\ \bibnamefont {Kheruntsyan}},\
  }\href {\doibase 10.1103/PhysRevLett.113.050601} {\bibfield  {journal}
  {\bibinfo  {journal} {Phys. Rev. Lett.}\ }\textbf {\bibinfo {volume} {113}},\
  \bibinfo {pages} {050601} (\bibinfo {year} {2014})}\BibitemShut {NoStop}%
\bibitem [{\citenamefont {{Andrei}}(2016)}]{integrablebose_Andrei2016}%
  \BibitemOpen
  \bibfield  {author} {\bibinfo {author} {\bibfnamefont {N.}~\bibnamefont
  {{Andrei}}},\ }\href@noop {} {\bibfield  {journal} {\bibinfo  {journal}
  {ArXiv e-prints}\ } (\bibinfo {year} {2016})},\ \Eprint
  {http://arxiv.org/abs/1606.08911} {arXiv:1606.08911 [cond-mat.quant-gas]}
  \BibitemShut {NoStop}%
\bibitem [{\citenamefont {Kormos}\ \emph {et~al.}(2014)\citenamefont {Kormos},
  \citenamefont {Collura},\ and\ \citenamefont
  {Calabrese}}]{1dbosepreth_kormos_2014}%
  \BibitemOpen
  \bibfield  {author} {\bibinfo {author} {\bibfnamefont {M.}~\bibnamefont
  {Kormos}}, \bibinfo {author} {\bibfnamefont {M.}~\bibnamefont {Collura}}, \
  and\ \bibinfo {author} {\bibfnamefont {P.}~\bibnamefont {Calabrese}},\ }\href
  {\doibase 10.1103/PhysRevA.89.013609} {\bibfield  {journal} {\bibinfo
  {journal} {Phys. Rev. A}\ }\textbf {\bibinfo {volume} {89}},\ \bibinfo
  {pages} {013609} (\bibinfo {year} {2014})}\BibitemShut {NoStop}%
\bibitem [{\citenamefont {Kormos}\ \emph {et~al.}(2013)\citenamefont {Kormos},
  \citenamefont {Shashi}, \citenamefont {Chou}, \citenamefont {Caux},\ and\
  \citenamefont {Imambekov}}]{1dbose_kormos_2013}%
  \BibitemOpen
  \bibfield  {author} {\bibinfo {author} {\bibfnamefont {M.}~\bibnamefont
  {Kormos}}, \bibinfo {author} {\bibfnamefont {A.}~\bibnamefont {Shashi}},
  \bibinfo {author} {\bibfnamefont {Y.-Z.}\ \bibnamefont {Chou}}, \bibinfo
  {author} {\bibfnamefont {J.-S.}\ \bibnamefont {Caux}}, \ and\ \bibinfo
  {author} {\bibfnamefont {A.}~\bibnamefont {Imambekov}},\ }\href {\doibase
  10.1103/PhysRevB.88.205131} {\bibfield  {journal} {\bibinfo  {journal} {Phys.
  Rev. B}\ }\textbf {\bibinfo {volume} {88}},\ \bibinfo {pages} {205131}
  (\bibinfo {year} {2013})}\BibitemShut {NoStop}%
\bibitem [{\citenamefont {Rigol}\ \emph {et~al.}(2008)\citenamefont {Rigol},
  \citenamefont {Dunjko},\ and\ \citenamefont {Olshanii}}]{Rigol_ETH}%
  \BibitemOpen
  \bibfield  {author} {\bibinfo {author} {\bibfnamefont {M.}~\bibnamefont
  {Rigol}}, \bibinfo {author} {\bibfnamefont {V.}~\bibnamefont {Dunjko}}, \
  and\ \bibinfo {author} {\bibfnamefont {M.}~\bibnamefont {Olshanii}},\ }\href
  {http://dx.doi.org/10.1038/nature06838} {\bibfield  {journal} {\bibinfo
  {journal} {Nature}\ }\textbf {\bibinfo {volume} {452}},\ \bibinfo {pages}
  {854 EP } (\bibinfo {year} {2008})}\BibitemShut {NoStop}%
\bibitem [{\citenamefont {Santos}\ and\ \citenamefont
  {Rigol}(2010{\natexlab{a}})}]{rigol_breakintegrability_thermali_2010}%
  \BibitemOpen
  \bibfield  {author} {\bibinfo {author} {\bibfnamefont {L.~F.}\ \bibnamefont
  {Santos}}\ and\ \bibinfo {author} {\bibfnamefont {M.}~\bibnamefont {Rigol}},\
  }\href {\doibase 10.1103/PhysRevE.82.031130} {\bibfield  {journal} {\bibinfo
  {journal} {Phys. Rev. E}\ }\textbf {\bibinfo {volume} {82}},\ \bibinfo
  {pages} {031130} (\bibinfo {year} {2010}{\natexlab{a}})}\BibitemShut
  {NoStop}%
\bibitem [{\citenamefont {Santos}\ and\ \citenamefont
  {Rigol}(2010{\natexlab{b}})}]{rigol_breakintegrability_2010}%
  \BibitemOpen
  \bibfield  {author} {\bibinfo {author} {\bibfnamefont {L.~F.}\ \bibnamefont
  {Santos}}\ and\ \bibinfo {author} {\bibfnamefont {M.}~\bibnamefont {Rigol}},\
  }\href {\doibase 10.1103/PhysRevE.81.036206} {\bibfield  {journal} {\bibinfo
  {journal} {Phys. Rev. E}\ }\textbf {\bibinfo {volume} {81}},\ \bibinfo
  {pages} {036206} (\bibinfo {year} {2010}{\natexlab{b}})}\BibitemShut
  {NoStop}%
\bibitem [{\citenamefont
  {Rigol}(2009{\natexlab{a}})}]{rigol_thermalization_2009}%
  \BibitemOpen
  \bibfield  {author} {\bibinfo {author} {\bibfnamefont {M.}~\bibnamefont
  {Rigol}},\ }\href {\doibase 10.1103/PhysRevA.80.053607} {\bibfield  {journal}
  {\bibinfo  {journal} {Phys. Rev. A}\ }\textbf {\bibinfo {volume} {80}},\
  \bibinfo {pages} {053607} (\bibinfo {year} {2009}{\natexlab{a}})}\BibitemShut
  {NoStop}%
\bibitem [{\citenamefont
  {Rigol}(2009{\natexlab{b}})}]{rigol_thermalization2_2009}%
  \BibitemOpen
  \bibfield  {author} {\bibinfo {author} {\bibfnamefont {M.}~\bibnamefont
  {Rigol}},\ }\href {\doibase 10.1103/PhysRevLett.103.100403} {\bibfield
  {journal} {\bibinfo  {journal} {Phys. Rev. Lett.}\ }\textbf {\bibinfo
  {volume} {103}},\ \bibinfo {pages} {100403} (\bibinfo {year}
  {2009}{\natexlab{b}})}\BibitemShut {NoStop}%
\bibitem [{\citenamefont {Biroli}\ \emph {et~al.}(2010)\citenamefont {Biroli},
  \citenamefont {Kollath},\ and\ \citenamefont
  {L\"auchli}}]{Biroli_thermalization_2010}%
  \BibitemOpen
  \bibfield  {author} {\bibinfo {author} {\bibfnamefont {G.}~\bibnamefont
  {Biroli}}, \bibinfo {author} {\bibfnamefont {C.}~\bibnamefont {Kollath}}, \
  and\ \bibinfo {author} {\bibfnamefont {A.~M.}\ \bibnamefont {L\"auchli}},\
  }\href {\doibase 10.1103/PhysRevLett.105.250401} {\bibfield  {journal}
  {\bibinfo  {journal} {Phys. Rev. Lett.}\ }\textbf {\bibinfo {volume} {105}},\
  \bibinfo {pages} {250401} (\bibinfo {year} {2010})}\BibitemShut {NoStop}%
\bibitem [{\citenamefont {Mori}\ \emph {et~al.}(2018)\citenamefont {Mori},
  \citenamefont {Ikeda}, \citenamefont {Kaminishi},\ and\ \citenamefont
  {Ueda}}]{thandpreth_2018}%
  \BibitemOpen
  \bibfield  {author} {\bibinfo {author} {\bibfnamefont {T.}~\bibnamefont
  {Mori}}, \bibinfo {author} {\bibfnamefont {T.~N.}\ \bibnamefont {Ikeda}},
  \bibinfo {author} {\bibfnamefont {E.}~\bibnamefont {Kaminishi}}, \ and\
  \bibinfo {author} {\bibfnamefont {M.}~\bibnamefont {Ueda}},\ }\href
  {http://stacks.iop.org/0953-4075/51/i=11/a=112001} {\bibfield  {journal}
  {\bibinfo  {journal} {Journal of Physics B: Atomic, Molecular and Optical
  Physics}\ }\textbf {\bibinfo {volume} {51}},\ \bibinfo {pages} {112001}
  (\bibinfo {year} {2018})}\BibitemShut {NoStop}%
\bibitem [{\citenamefont {Mallayya}\ and\ \citenamefont
  {Rigol}(2018)}]{rigol_thermalization_2018}%
  \BibitemOpen
  \bibfield  {author} {\bibinfo {author} {\bibfnamefont {K.}~\bibnamefont
  {Mallayya}}\ and\ \bibinfo {author} {\bibfnamefont {M.}~\bibnamefont
  {Rigol}},\ }\href {\doibase 10.1103/PhysRevLett.120.070603} {\bibfield
  {journal} {\bibinfo  {journal} {Phys. Rev. Lett.}\ }\textbf {\bibinfo
  {volume} {120}},\ \bibinfo {pages} {070603} (\bibinfo {year}
  {2018})}\BibitemShut {NoStop}%
\bibitem [{\citenamefont {Tang}\ \emph {et~al.}(2018)\citenamefont {Tang},
  \citenamefont {Kao}, \citenamefont {Li}, \citenamefont {Seo}, \citenamefont
  {Mallayya}, \citenamefont {Rigol}, \citenamefont {Gopalakrishnan},\ and\
  \citenamefont {Lev}}]{rigol_thermalization_NC_2018}%
  \BibitemOpen
  \bibfield  {author} {\bibinfo {author} {\bibfnamefont {Y.}~\bibnamefont
  {Tang}}, \bibinfo {author} {\bibfnamefont {W.}~\bibnamefont {Kao}}, \bibinfo
  {author} {\bibfnamefont {K.-Y.}\ \bibnamefont {Li}}, \bibinfo {author}
  {\bibfnamefont {S.}~\bibnamefont {Seo}}, \bibinfo {author} {\bibfnamefont
  {K.}~\bibnamefont {Mallayya}}, \bibinfo {author} {\bibfnamefont
  {M.}~\bibnamefont {Rigol}}, \bibinfo {author} {\bibfnamefont
  {S.}~\bibnamefont {Gopalakrishnan}}, \ and\ \bibinfo {author} {\bibfnamefont
  {B.~L.}\ \bibnamefont {Lev}},\ }\href {\doibase 10.1103/PhysRevX.8.021030}
  {\bibfield  {journal} {\bibinfo  {journal} {Phys. Rev. X}\ }\textbf {\bibinfo
  {volume} {8}},\ \bibinfo {pages} {021030} (\bibinfo {year}
  {2018})}\BibitemShut {NoStop}%
\bibitem [{\citenamefont {Eckstein}\ \emph {et~al.}(2010)\citenamefont
  {Eckstein}, \citenamefont {Kollar},\ and\ \citenamefont
  {Werner}}]{eckstein_hubbard_2010}%
  \BibitemOpen
  \bibfield  {author} {\bibinfo {author} {\bibfnamefont {M.}~\bibnamefont
  {Eckstein}}, \bibinfo {author} {\bibfnamefont {M.}~\bibnamefont {Kollar}}, \
  and\ \bibinfo {author} {\bibfnamefont {P.}~\bibnamefont {Werner}},\ }\href
  {\doibase 10.1103/PhysRevB.81.115131} {\bibfield  {journal} {\bibinfo
  {journal} {Phys. Rev. B}\ }\textbf {\bibinfo {volume} {81}},\ \bibinfo
  {pages} {115131} (\bibinfo {year} {2010})}\BibitemShut {NoStop}%
\bibitem [{\citenamefont {{Van Regemortel}}\ \emph {et~al.}(2018)\citenamefont
  {{Van Regemortel}}, \citenamefont {{Kurkjian}}, \citenamefont {{Carusotto}},\
  and\ \citenamefont {{Wouters}}}]{boseprethramp2018}%
  \BibitemOpen
  \bibfield  {author} {\bibinfo {author} {\bibfnamefont {M.}~\bibnamefont {{Van
  Regemortel}}}, \bibinfo {author} {\bibfnamefont {H.}~\bibnamefont
  {{Kurkjian}}}, \bibinfo {author} {\bibfnamefont {I.}~\bibnamefont
  {{Carusotto}}}, \ and\ \bibinfo {author} {\bibfnamefont {M.}~\bibnamefont
  {{Wouters}}},\ }\href@noop {} {\bibfield  {journal} {\bibinfo  {journal}
  {ArXiv e-prints}\ } (\bibinfo {year} {2018})},\ \Eprint
  {http://arxiv.org/abs/1803.07459} {arXiv:1803.07459 [cond-mat.quant-gas]}
  \BibitemShut {NoStop}%
\bibitem [{\citenamefont {Barnett}\ \emph {et~al.}(2011)\citenamefont
  {Barnett}, \citenamefont {Polkovnikov},\ and\ \citenamefont
  {Vengalattore}}]{preth_spin1_2011}%
  \BibitemOpen
  \bibfield  {author} {\bibinfo {author} {\bibfnamefont {R.}~\bibnamefont
  {Barnett}}, \bibinfo {author} {\bibfnamefont {A.}~\bibnamefont
  {Polkovnikov}}, \ and\ \bibinfo {author} {\bibfnamefont {M.}~\bibnamefont
  {Vengalattore}},\ }\href {\doibase 10.1103/PhysRevA.84.023606} {\bibfield
  {journal} {\bibinfo  {journal} {Phys. Rev. A}\ }\textbf {\bibinfo {volume}
  {84}},\ \bibinfo {pages} {023606} (\bibinfo {year} {2011})}\BibitemShut
  {NoStop}%
\bibitem [{\citenamefont {Cosme}(2018)}]{bose_prerh_cosme_2018}%
  \BibitemOpen
  \bibfield  {author} {\bibinfo {author} {\bibfnamefont {J.~G.}\ \bibnamefont
  {Cosme}},\ }\href {\doibase 10.1103/PhysRevA.97.043610} {\bibfield  {journal}
  {\bibinfo  {journal} {Phys. Rev. A}\ }\textbf {\bibinfo {volume} {97}},\
  \bibinfo {pages} {043610} (\bibinfo {year} {2018})}\BibitemShut {NoStop}%
\bibitem [{\citenamefont {{Neyenhuis}}\ \emph {et~al.}(2016)\citenamefont
  {{Neyenhuis}}, \citenamefont {{Smith}}, \citenamefont {{Lee}}, \citenamefont
  {{Zhang}}, \citenamefont {{Richerme}}, \citenamefont {{Hess}}, \citenamefont
  {{Gong}}, \citenamefont {{Gorshkov}},\ and\ \citenamefont
  {{Monroe}}}]{prethexp_Neye2016}%
  \BibitemOpen
  \bibfield  {author} {\bibinfo {author} {\bibfnamefont {B.}~\bibnamefont
  {{Neyenhuis}}}, \bibinfo {author} {\bibfnamefont {J.}~\bibnamefont
  {{Smith}}}, \bibinfo {author} {\bibfnamefont {A.~C.}\ \bibnamefont {{Lee}}},
  \bibinfo {author} {\bibfnamefont {J.}~\bibnamefont {{Zhang}}}, \bibinfo
  {author} {\bibfnamefont {P.}~\bibnamefont {{Richerme}}}, \bibinfo {author}
  {\bibfnamefont {P.~W.}\ \bibnamefont {{Hess}}}, \bibinfo {author}
  {\bibfnamefont {Z.-X.}\ \bibnamefont {{Gong}}}, \bibinfo {author}
  {\bibfnamefont {A.~V.}\ \bibnamefont {{Gorshkov}}}, \ and\ \bibinfo {author}
  {\bibfnamefont {C.}~\bibnamefont {{Monroe}}},\ }\href@noop {} {\bibfield
  {journal} {\bibinfo  {journal} {ArXiv e-prints}\ } (\bibinfo {year}
  {2016})},\ \Eprint {http://arxiv.org/abs/1608.00681} {arXiv:1608.00681
  [quant-ph]} \BibitemShut {NoStop}%
\bibitem [{\citenamefont {Gring}\ \emph {et~al.}(2012)\citenamefont {Gring},
  \citenamefont {Kuhnert}, \citenamefont {Langen}, \citenamefont {Kitagawa},
  \citenamefont {Rauer}, \citenamefont {Schreitl}, \citenamefont {Mazets},
  \citenamefont {Smith}, \citenamefont {Demler},\ and\ \citenamefont
  {Schmiedmayer}}]{bose_exp_Gring1318}%
  \BibitemOpen
  \bibfield  {author} {\bibinfo {author} {\bibfnamefont {M.}~\bibnamefont
  {Gring}}, \bibinfo {author} {\bibfnamefont {M.}~\bibnamefont {Kuhnert}},
  \bibinfo {author} {\bibfnamefont {T.}~\bibnamefont {Langen}}, \bibinfo
  {author} {\bibfnamefont {T.}~\bibnamefont {Kitagawa}}, \bibinfo {author}
  {\bibfnamefont {B.}~\bibnamefont {Rauer}}, \bibinfo {author} {\bibfnamefont
  {M.}~\bibnamefont {Schreitl}}, \bibinfo {author} {\bibfnamefont
  {I.}~\bibnamefont {Mazets}}, \bibinfo {author} {\bibfnamefont {D.~A.}\
  \bibnamefont {Smith}}, \bibinfo {author} {\bibfnamefont {E.}~\bibnamefont
  {Demler}}, \ and\ \bibinfo {author} {\bibfnamefont {J.}~\bibnamefont
  {Schmiedmayer}},\ }\href {\doibase 10.1126/science.1224953} {\bibfield
  {journal} {\bibinfo  {journal} {Science}\ }\textbf {\bibinfo {volume}
  {337}},\ \bibinfo {pages} {1318} (\bibinfo {year} {2012})}\BibitemShut
  {NoStop}%
\bibitem [{\citenamefont {{Eigen}}\ \emph {et~al.}(2018)\citenamefont
  {{Eigen}}, \citenamefont {{Glidden}}, \citenamefont {{Lopes}}, \citenamefont
  {{Cornell}}, \citenamefont {{Smith}},\ and\ \citenamefont
  {{Hadzibabic}}}]{bose_prethexp_2018}%
  \BibitemOpen
  \bibfield  {author} {\bibinfo {author} {\bibfnamefont {C.}~\bibnamefont
  {{Eigen}}}, \bibinfo {author} {\bibfnamefont {J.~A.~P.}\ \bibnamefont
  {{Glidden}}}, \bibinfo {author} {\bibfnamefont {R.}~\bibnamefont {{Lopes}}},
  \bibinfo {author} {\bibfnamefont {E.~A.}\ \bibnamefont {{Cornell}}}, \bibinfo
  {author} {\bibfnamefont {R.~P.}\ \bibnamefont {{Smith}}}, \ and\ \bibinfo
  {author} {\bibfnamefont {Z.}~\bibnamefont {{Hadzibabic}}},\ }\href@noop {}
  {\bibfield  {journal} {\bibinfo  {journal} {ArXiv e-prints}\ } (\bibinfo
  {year} {2018})},\ \Eprint {http://arxiv.org/abs/1805.09802} {arXiv:1805.09802
  [cond-mat.quant-gas]} \BibitemShut {NoStop}%
\bibitem [{\citenamefont {{Stark}}\ and\ \citenamefont
  {{Kollar}}(2013)}]{kollarperturbation_2013}%
  \BibitemOpen
  \bibfield  {author} {\bibinfo {author} {\bibfnamefont {M.}~\bibnamefont
  {{Stark}}}\ and\ \bibinfo {author} {\bibfnamefont {M.}~\bibnamefont
  {{Kollar}}},\ }\href@noop {} {\bibfield  {journal} {\bibinfo  {journal}
  {ArXiv e-prints}\ } (\bibinfo {year} {2013})},\ \Eprint
  {http://arxiv.org/abs/1308.1610} {arXiv:1308.1610 [cond-mat.str-el]}
  \BibitemShut {NoStop}%
\bibitem [{\citenamefont {Chiocchetta}\ \emph {et~al.}(2016)\citenamefont
  {Chiocchetta}, \citenamefont {Tavora}, \citenamefont {Gambassi},\ and\
  \citenamefont {Mitra}}]{perturb_prethandth_2016}%
  \BibitemOpen
  \bibfield  {author} {\bibinfo {author} {\bibfnamefont {A.}~\bibnamefont
  {Chiocchetta}}, \bibinfo {author} {\bibfnamefont {M.}~\bibnamefont {Tavora}},
  \bibinfo {author} {\bibfnamefont {A.}~\bibnamefont {Gambassi}}, \ and\
  \bibinfo {author} {\bibfnamefont {A.}~\bibnamefont {Mitra}},\ }\href
  {\doibase 10.1103/PhysRevB.94.134311} {\bibfield  {journal} {\bibinfo
  {journal} {Phys. Rev. B}\ }\textbf {\bibinfo {volume} {94}},\ \bibinfo
  {pages} {134311} (\bibinfo {year} {2016})}\BibitemShut {NoStop}%
\bibitem [{\citenamefont {Huang}\ and\ \citenamefont
  {Cazalilla}(2019)}]{Huang2019}%
  \BibitemOpen
  \bibfield  {author} {\bibinfo {author} {\bibfnamefont {C.-H.}\ \bibnamefont
  {Huang}}\ and\ \bibinfo {author} {\bibfnamefont {M.~A.}\ \bibnamefont
  {Cazalilla}},\ }\href {\doibase 10.1103/PhysRevA.99.063612} {\bibfield
  {journal} {\bibinfo  {journal} {Phys. Rev. A}\ }\textbf {\bibinfo {volume}
  {99}},\ \bibinfo {pages} {063612} (\bibinfo {year} {2019})}\BibitemShut
  {NoStop}%
\bibitem [{\citenamefont {Yip}\ and\ \citenamefont {Ho}(1999)}]{Yip1998}%
  \BibitemOpen
  \bibfield  {author} {\bibinfo {author} {\bibfnamefont {S.-K.}\ \bibnamefont
  {Yip}}\ and\ \bibinfo {author} {\bibfnamefont {T.-L.}\ \bibnamefont {Ho}},\
  }\href {\doibase 10.1103/PhysRevA.59.4653} {\bibfield  {journal} {\bibinfo
  {journal} {Phys. Rev. A}\ }\textbf {\bibinfo {volume} {59}},\ \bibinfo
  {pages} {4653} (\bibinfo {year} {1999})}\BibitemShut {NoStop}%
\bibitem [{\citenamefont {Blatt}\ \emph {et~al.}(2011)\citenamefont {Blatt},
  \citenamefont {Nicholson}, \citenamefont {Bloom}, \citenamefont {Williams},
  \citenamefont {Thomsen}, \citenamefont {Julienne},\ and\ \citenamefont
  {Ye}}]{Blatt2011}%
  \BibitemOpen
  \bibfield  {author} {\bibinfo {author} {\bibfnamefont {S.}~\bibnamefont
  {Blatt}}, \bibinfo {author} {\bibfnamefont {T.~L.}\ \bibnamefont
  {Nicholson}}, \bibinfo {author} {\bibfnamefont {B.~J.}\ \bibnamefont
  {Bloom}}, \bibinfo {author} {\bibfnamefont {J.~R.}\ \bibnamefont {Williams}},
  \bibinfo {author} {\bibfnamefont {J.~W.}\ \bibnamefont {Thomsen}}, \bibinfo
  {author} {\bibfnamefont {P.~S.}\ \bibnamefont {Julienne}}, \ and\ \bibinfo
  {author} {\bibfnamefont {J.}~\bibnamefont {Ye}},\ }\href {\doibase
  10.1103/PhysRevLett.107.073202} {\bibfield  {journal} {\bibinfo  {journal}
  {Phys. Rev. Lett.}\ }\textbf {\bibinfo {volume} {107}},\ \bibinfo {pages}
  {073202} (\bibinfo {year} {2011})}\BibitemShut {NoStop}%
\bibitem [{\citenamefont {Ciury\l{}o}\ \emph {et~al.}(2005)\citenamefont
  {Ciury\l{}o}, \citenamefont {Tiesinga},\ and\ \citenamefont
  {Julienne}}]{Ciurylo2005}%
  \BibitemOpen
  \bibfield  {author} {\bibinfo {author} {\bibfnamefont {R.}~\bibnamefont
  {Ciury\l{}o}}, \bibinfo {author} {\bibfnamefont {E.}~\bibnamefont
  {Tiesinga}}, \ and\ \bibinfo {author} {\bibfnamefont {P.~S.}\ \bibnamefont
  {Julienne}},\ }\href {\doibase 10.1103/PhysRevA.71.030701} {\bibfield
  {journal} {\bibinfo  {journal} {Phys. Rev. A}\ }\textbf {\bibinfo {volume}
  {71}},\ \bibinfo {pages} {030701} (\bibinfo {year} {2005})}\BibitemShut
  {NoStop}%
\bibitem [{\citenamefont {Enomoto}\ \emph {et~al.}(2008)\citenamefont
  {Enomoto}, \citenamefont {Kasa}, \citenamefont {Kitagawa},\ and\
  \citenamefont {Takahashi}}]{Enomoto2008}%
  \BibitemOpen
  \bibfield  {author} {\bibinfo {author} {\bibfnamefont {K.}~\bibnamefont
  {Enomoto}}, \bibinfo {author} {\bibfnamefont {K.}~\bibnamefont {Kasa}},
  \bibinfo {author} {\bibfnamefont {M.}~\bibnamefont {Kitagawa}}, \ and\
  \bibinfo {author} {\bibfnamefont {Y.}~\bibnamefont {Takahashi}},\ }\href
  {\doibase 10.1103/PhysRevLett.101.203201} {\bibfield  {journal} {\bibinfo
  {journal} {Phys. Rev. Lett.}\ }\textbf {\bibinfo {volume} {101}},\ \bibinfo
  {pages} {203201} (\bibinfo {year} {2008})}\BibitemShut {NoStop}%
\bibitem [{\citenamefont {Yan}\ \emph {et~al.}(2013)\citenamefont {Yan},
  \citenamefont {DeSalvo}, \citenamefont {Ramachandhran}, \citenamefont {Pu},\
  and\ \citenamefont {Killian}}]{Yan2010}%
  \BibitemOpen
  \bibfield  {author} {\bibinfo {author} {\bibfnamefont {M.}~\bibnamefont
  {Yan}}, \bibinfo {author} {\bibfnamefont {B.~J.}\ \bibnamefont {DeSalvo}},
  \bibinfo {author} {\bibfnamefont {B.}~\bibnamefont {Ramachandhran}}, \bibinfo
  {author} {\bibfnamefont {H.}~\bibnamefont {Pu}}, \ and\ \bibinfo {author}
  {\bibfnamefont {T.~C.}\ \bibnamefont {Killian}},\ }\href {\doibase
  10.1103/PhysRevLett.110.123201} {\bibfield  {journal} {\bibinfo  {journal}
  {Phys. Rev. Lett.}\ }\textbf {\bibinfo {volume} {110}},\ \bibinfo {pages}
  {123201} (\bibinfo {year} {2013})}\BibitemShut {NoStop}%
\bibitem [{\citenamefont {Fedichev}\ \emph {et~al.}(1996)\citenamefont
  {Fedichev}, \citenamefont {Kagan}, \citenamefont {Shlyapnikov},\ and\
  \citenamefont {Walraven}}]{PhysRevLett.77.2913_walraven}%
  \BibitemOpen
  \bibfield  {author} {\bibinfo {author} {\bibfnamefont {P.~O.}\ \bibnamefont
  {Fedichev}}, \bibinfo {author} {\bibfnamefont {Y.}~\bibnamefont {Kagan}},
  \bibinfo {author} {\bibfnamefont {G.~V.}\ \bibnamefont {Shlyapnikov}}, \ and\
  \bibinfo {author} {\bibfnamefont {J.~T.~M.}\ \bibnamefont {Walraven}},\
  }\href {\doibase 10.1103/PhysRevLett.77.2913} {\bibfield  {journal} {\bibinfo
   {journal} {Phys. Rev. Lett.}\ }\textbf {\bibinfo {volume} {77}},\ \bibinfo
  {pages} {2913} (\bibinfo {year} {1996})}\BibitemShut {NoStop}%
\bibitem [{\citenamefont {Theis}\ \emph {et~al.}(2004)\citenamefont {Theis},
  \citenamefont {Thalhammer}, \citenamefont {Winkler}, \citenamefont {Hellwig},
  \citenamefont {Ruff}, \citenamefont {Grimm},\ and\ \citenamefont
  {Denschlag}}]{PhysRevLett.93.123001_hecker}%
  \BibitemOpen
  \bibfield  {author} {\bibinfo {author} {\bibfnamefont {M.}~\bibnamefont
  {Theis}}, \bibinfo {author} {\bibfnamefont {G.}~\bibnamefont {Thalhammer}},
  \bibinfo {author} {\bibfnamefont {K.}~\bibnamefont {Winkler}}, \bibinfo
  {author} {\bibfnamefont {M.}~\bibnamefont {Hellwig}}, \bibinfo {author}
  {\bibfnamefont {G.}~\bibnamefont {Ruff}}, \bibinfo {author} {\bibfnamefont
  {R.}~\bibnamefont {Grimm}}, \ and\ \bibinfo {author} {\bibfnamefont {J.~H.}\
  \bibnamefont {Denschlag}},\ }\href {\doibase 10.1103/PhysRevLett.93.123001}
  {\bibfield  {journal} {\bibinfo  {journal} {Phys. Rev. Lett.}\ }\textbf
  {\bibinfo {volume} {93}},\ \bibinfo {pages} {123001} (\bibinfo {year}
  {2004})}\BibitemShut {NoStop}%
\bibitem [{\citenamefont {Yamazaki}\ \emph
  {et~al.}(2010{\natexlab{b}})\citenamefont {Yamazaki}, \citenamefont {Taie},
  \citenamefont {Sugawa},\ and\ \citenamefont
  {Takahashi}}]{PhysRevLett.105.050405_takahashi}%
  \BibitemOpen
  \bibfield  {author} {\bibinfo {author} {\bibfnamefont {R.}~\bibnamefont
  {Yamazaki}}, \bibinfo {author} {\bibfnamefont {S.}~\bibnamefont {Taie}},
  \bibinfo {author} {\bibfnamefont {S.}~\bibnamefont {Sugawa}}, \ and\ \bibinfo
  {author} {\bibfnamefont {Y.}~\bibnamefont {Takahashi}},\ }\href {\doibase
  10.1103/PhysRevLett.105.050405} {\bibfield  {journal} {\bibinfo  {journal}
  {Phys. Rev. Lett.}\ }\textbf {\bibinfo {volume} {105}},\ \bibinfo {pages}
  {050405} (\bibinfo {year} {2010}{\natexlab{b}})}\BibitemShut {NoStop}%
\bibitem [{\citenamefont {Gramsch}\ and\ \citenamefont
  {Potthoff}(2015)}]{PhysRevB.92.235135}%
  \BibitemOpen
  \bibfield  {author} {\bibinfo {author} {\bibfnamefont {C.}~\bibnamefont
  {Gramsch}}\ and\ \bibinfo {author} {\bibfnamefont {M.}~\bibnamefont
  {Potthoff}},\ }\href {\doibase 10.1103/PhysRevB.92.235135} {\bibfield
  {journal} {\bibinfo  {journal} {Phys. Rev. B}\ }\textbf {\bibinfo {volume}
  {92}},\ \bibinfo {pages} {235135} (\bibinfo {year} {2015})}\BibitemShut
  {NoStop}%
\bibitem [{\citenamefont {{Tsuji}}\ and\ \citenamefont
  {{Werner}}(2013)}]{perturb_werner2013}%
  \BibitemOpen
  \bibfield  {author} {\bibinfo {author} {\bibfnamefont {N.}~\bibnamefont
  {{Tsuji}}}\ and\ \bibinfo {author} {\bibfnamefont {P.}~\bibnamefont
  {{Werner}}},\ }\href {\doibase 10.1103/PhysRevB.88.165115} {\bibfield
  {journal} {\bibinfo  {journal} {\prb}\ }\textbf {\bibinfo {volume} {88}},\
  \bibinfo {eid} {165115} (\bibinfo {year} {2013})},\ \Eprint
  {http://arxiv.org/abs/1306.0307} {arXiv:1306.0307 [cond-mat.str-el]}
  \BibitemShut {NoStop}%
\end{thebibliography}%

\end{document}